\documentclass[3p,,preprint,12pt]{elsarticle}
\pdfoutput=1
\makeatletter\if@twocolumn\PassOptionsToPackage{switch}{lineno}\else\fi\makeatother

\usepackage{geometry}
\usepackage{subfiles}

\usepackage{lmodern}
\usepackage[english]{babel}
\usepackage{microtype}

\usepackage{tabulary,xcolor}
\usepackage{amsfonts,amsmath,amssymb,amsthm}
\usepackage[T1]{fontenc}
\makeatletter
\let\save@ps@pprintTitle\ps@pprintTitle
\def\ps@pprintTitle{\save@ps@pprintTitle\gdef\@oddfoot{\footnotesize\itshape \null\hfill\today}}
\def\hlinewd#1{%
  \noalign{\ifnum0=`}\fi\hrule \@height #1%
  \futurelet\reserved@a\@xhline}

\AtBeginDocument{\ifNAT@numbers \biboptions{sort&compress}\fi}
\makeatother

  \renewenvironment{abstract}{\global\setbox\absbox=\vbox\bgroup
    \hsize=\textwidth%
  \noindent\unskip\textbf{}
   \par\medskip\noindent\unskip\ignorespaces}
   {\egroup}
  
\usepackage{ifluatex}
\ifluatex
\usepackage{fontspec}
\defaultfontfeatures{Ligatures=TeX}
\usepackage[]{unicode-math}
\unimathsetup{math-style=TeX}
\else 
\usepackage[utf8]{inputenc}
\fi 
\ifluatex\else\usepackage{stmaryrd}\fi

\usepackage[acronym]{glossaries}

\usepackage{booktabs}
\usepackage{enumitem}
  
\usepackage{url,multirow,morefloats,floatflt,cancel,tfrupee}
\makeatletter

\AtBeginDocument{\@ifpackageloaded{textcomp}{}{\usepackage{textcomp}}}
\makeatother
\usepackage{colortbl}
\usepackage{xcolor}
\usepackage{pifont}
\usepackage[nointegrals]{wasysym}
\makeatletter

\def\mcWidth#1{\csname TY@F#1\endcsname+\tabcolsep}

\def\cAlignHack{\rightskip\@flushglue\leftskip\@flushglue\parindent\z@\parfillskip\z@skip}
\def\rAlignHack{\rightskip\z@skip\leftskip\@flushglue \parindent\z@\parfillskip\z@skip}

\@ifundefined{etal}{}{}

\usepackage{ifxetex}
\ifxetex\else\if@twocolumn\@ifpackageloaded{stfloats}{}{\usepackage{dblfloatfix}}\fi\fi

\AtBeginDocument{
\expandafter\ifx\csname eqalign\endcsname\relax
\def\eqalign#1{\null\vcenter{\def\\{\cr}\openup\jot\m@th
  \ialign{\strut$\displaystyle{##}$\hfil&$\displaystyle{{}##}$\hfil
      \crcr#1\crcr}}\,}
\fi
}

\AtBeginDocument{%
  \@ifpackageloaded{endfloat}%
   {\renewcommand\efloat@iwrite[1]{\immediate\expandafter\protected@write\csname efloat@post#1\endcsname{}}}{\newif\ifefloat@tables}%
}%

\def\BreakURLText#1{\@tfor\brk@tempa:=#1\do{\brk@tempa\hskip0pt}}
\let\lt=<
\let\gt=>
\def\processVert{\ifmmode|\else\textbar\fi}

\@ifundefined{subparagraph}{
\def\subparagraph{\@startsection{paragraph}{5}{2\parindent}{0ex plus 0.1ex minus 0.1ex}%
{0ex}{\normalfont\small\itshape}}%
}{}

\newcommand\role[1]{\unskip}
\newcommand\aucollab[1]{\unskip}
  
\@ifundefined{tsGraphicsScaleX}{\gdef\tsGraphicsScaleX{1}}{}
\@ifundefined{tsGraphicsScaleY}{\gdef\tsGraphicsScaleY{.9}}{}
\def\checkGraphicsWidth{\ifdim\Gin@nat@width>\linewidth
	\tsGraphicsScaleX\linewidth\else\Gin@nat@width\fi}

\def\checkGraphicsHeight{\ifdim\Gin@nat@height>.9\textheight
	\tsGraphicsScaleY\textheight\else\Gin@nat@height\fi}

\def\fixFloatSize#1{}
\let\ts@includegraphics\includegraphics

\def\inlinegraphic[#1]#2{{\edef\@tempa{#1}\edef\baseline@shift{\ifx\@tempa\@empty0\else#1\fi}\edef\tempZ{\the\numexpr(\numexpr(\baseline@shift*\f@size/100))}\protect\raisebox{\tempZ pt}{\ts@includegraphics{#2}}}}

\AtBeginDocument{\def\includegraphics{\@ifnextchar[{\ts@includegraphics}{\ts@includegraphics[width=\checkGraphicsWidth,height=\checkGraphicsHeight,keepaspectratio]}}}

\DeclareMathAlphabet{\mathpzc}{OT1}{pzc}{m}{it}

\def\URL#1#2{\@ifundefined{href}{#2}{\href{#1}{#2}}}

\def\UrlOrds{\do\*\do\-\do\~\do\'\do\"\do\-}%
\g@addto@macro{\UrlBreaks}{\UrlOrds}

\edef\fntEncoding{\f@encoding}

\makeatother

\newif\ifmultipleabstract\multipleabstractfalse%
\usepackage[ruled,vlined]{algorithm2e}
\SetKwInput{KwHyper}{Hyperparameter}
\usepackage{verbatim}
\usepackage{listings}

\usepackage{graphicx, graphics}
\usepackage{caption}
\usepackage{subcaption}
\usepackage{pgf,tikz}
\usetikzlibrary{arrows,automata,decorations.pathreplacing}

\usepackage{standalone}

\usepackage{cleveref}

\newcommand{\bB}{\mathbf{B}}

\newcommand{\bI}{\mathbf{I}}

\newcommand{\bGamma}{\mathbf{\Gamma}}

\newcommand{\bSigma}{\mathbf{\Sigma}}

\newcommand{\bLambda}{\mathbf{\Lambda}}

\newcommand{\bbeta}{\boldsymbol{\beta}}

\newcommand{\bepsilon}{\boldsymbol{\epsilon}}

\newcommand{\bOmega}{\boldsymbol{\Omega}}

\newcommand{\bx}{\mathbf{x}}

\newcommand{\by}{\boldsymbol{y}}
\newcommand{\bY}{\mathbf{Y}}
\newcommand{\bz}{\boldsymbol{z}}

\newcommand{\bF}{\mathbf{F}}

\newcommand{\0}{\mathbf{0}}

\newacronym{bicnet}{BICNet}{\textbf{B}ayesian \textbf{I}ntrinsic \textbf{C}onnectivity \textbf{Net}work}
\newacronym{gift}{GIFT}{Group ICA Of fMRI Toolbox}
\newacronym{icn}{ICN}{intrinsic connectivity network}
\newacronym{ica}{ICA}{independent component analysis}
\newacronym{fa}{FA}{factor analysis}
\newacronym{pca}{PCA}{principal component analysis}
\newacronym{asis}{ASIS}{ancillary-sufficiency interweaving strategy}

\newacronym{fmri}{fMRI}{functional magnetic resonance imaging}
\newacronym{bold}{BOLD}{blood-oxygen-level-dependent}
\newacronym{rsfmri}{rfMRI}{resting-state functional magnetic resonance imaging}
\newacronym{tfmri}{tfMRI}{task-related functional magnetic resonance imaging}
\newacronym{dag}{DAG}{directed acyclic graph}
\newacronym{sv}{SV}{stochastic volatility}
\newacronym{aal}{AAL}{automated anatomical labeling}
\newacronym{roi}{ROI}{region of interest}
\newacronym{ar}{AR}{autoregressive}
\newacronym{hcp}{HCP}{Human Connectome Project}
\newacronym{aic}{AIC}{Akaike information criterion}
\newacronym{bic}{BIC}{Bayesian Information Criterion}
\newacronym{dic}{DIC}{Deviance information criterion}
\newacronym{ks}{KS}{Kolmogorov-Smirnov}
\newacronym{mae}{MAE}{mean absolute error}
\newacronym{rmse}{RMSE}{root-mean-square error}

\newacronym{mfg}{MFG}{middle frontal gyrus}
\newacronym{itg}{ITG}{inferior temporal gyrus}
\newacronym{ifg}{IFG}{inferior frontal gyrus}
\newacronym{pcg}{PCG}{posterior cingulate gyrus}
\newacronym{precg}{PreCG}{precental gyrus}
\newacronym{hip}{HIP}{hippocampus}
\newacronym{put}{PUT}{putamen}
\newacronym{orbmid}{ORBmid}{orbital part of middle frontal gyrus}
\newacronym{ifgoperc}{IFGoperc}{opercular part of inferior frontal gyrus}
\newacronym{stg}{STG}{superior temporal gyrus}
\newacronym{sfg}{SFG}{superior frontal gyrus}
\newacronym{sma}{SMA}{supplementary motor area}
\newacronym{pcun}{PCUN}{precuneus}
\newacronym{ang}{ANG}{angular gyrus}
\newacronym{amyg}{AMYG}{amygdala}
\newacronym{cau}{CAU}{caudate nucleus}

\begin{document}

\begin{abstract}
\Glspl{icn} are specific dynamic functional brain networks that are consistently found under various conditions including rest and task. Studies have shown that some stimuli actually activate intrinsic connectivity through either suppression, excitation, moderation or modification. Nevertheless, the structure of \glspl{icn} and task-related effects on \glspl{icn} are not yet fully understood. In this paper, we propose a \gls{bicnet} model to identify the \glspl{icn} and quantify the task-related effects on the \gls{icn} dynamics. Using an extended Bayesian dynamic sparse latent factor model, the proposed \gls{bicnet} has the following advantages: (1) it simultaneously identifies the individual \glspl{icn} and group-level \gls{icn} spatial maps; (2) it robustly identifies \glspl{icn} by jointly modeling \gls{rsfmri} and \gls{tfmri}; (3) compared to \gls{ica}-based methods, it can quantify the difference of \glspl{icn} amplitudes across different states; (4) it automatically performs feature selection through the sparsity of the \glspl{icn} rather than ad-hoc thresholding. The proposed \gls{bicnet} was applied to the \gls{rsfmri} and language \gls{tfmri} data from the \gls{hcp} and the analysis identified several \glspl{icn} related to distinct language processing functions.
\end{abstract}


\begin{frontmatter}

\title{
    BICNet: A Bayesian Approach for Estimating Task Effects\newline on Intrinsic Connectivity Networks in fMRI Data
}

\author[kaust]{Meini Tang}
\author[malay]{Chee-Ming Ting}
\author[kaust]{Hernando Ombao\corref{cor1}}
\ead{hernando.ombao@kaust.edu.sa}
\cortext[cor1]{Corresponding author}
\address[kaust]{King Abdullah University of Science and Technology, Thuwal, 23955 Saudi Arabia}
\address[malay]{Monash University Malaysia, Subang Jaya, 47500 Malaysia}

\date{Received: date / Accepted: date}
    
\end{frontmatter}

\section{Introduction}
Functional brain networks are dynamically evolving to satisfy our ongoing demands, resulting from either external stimuli or internal processes. However, many studies have reported high spatial similarity and relatively low spatial dissimilarity between dynamic functional brain networks under resting and task-related states \citep{Smith2009a,Gordon2012,Cole2014a}. Also, some studies observe some networks consistently presenting across populations, including both healthy subjects and patients. It indicates that there is an \acrfull{icn} that persists under various cognitive conditions and across a population. These observations suggest that dynamic functional brain networks consist of both on-demand dynamics and intrinsic dynamics that support the fundamental brain functions. Recent research demonstrates that intrinsic connectivity correlates with cognition and behavioral performance \citep{Greicius2008}. Some studies show evidence of alterations in \gls{icn} in patients with neurological disorders, e.g., Alzheimer's disease, epilepsy, and schizophrenia, relative to the healthy control population. These findings strongly indicate that intrinsic connectivity can be a biomarker of neurological disorders \citep{Greicius2004,Greicius2008,Broyd2009,Goodkind2015}.

This paper aims to identify dynamic \glspl{icn} at the individual level and the corresponding intrinsic spatial maps at the group level using the proposed \acrfull{bicnet} model. Moreover, we aim to develop a formal statistical framework under which neuroscientists can compare the dynamics of different \glspl{icn} under different cognitive conditions, such as resting vs. task-related states. Note that we use connectivity and network interchangeably in this paper.
 
The key question is how to define the concept of \gls{icn}. Early studies consider resting-state connectivity networks as \glspl{icn} \citep{Fox2005,Seeley2007,Doll2015}. However, resting-state \glspl{icn} are confounded by unconscious context \citep{Laird2011} and hard to be related to certain cognitive functions. In the proposed \gls{bicnet} model, we pool the information from both \gls{rsfmri} and \gls{tfmri} together to improve robustness and interpretability. Then the question turns into in what form \gls{icn} exists in the dynamic functional networks under resting and task-related states. We adopt an intuitive assumption that some brain regions are intrinsically linked to networks with specialized roles in information processing. Then, a cognitive task dynamically engages multiple intrinsic networks \citep{Gratton2018}. In other words, \glspl{icn} are considered as unobservable components of dynamic functional networks that have relatively stable spatial structures but dynamic activities. The stimuli can activate \glspl{icn} through either suppression, excitation, moderation, or modification. This definition can be naturally translated into a latent factor model, $\by_t = \bLambda \boldsymbol{f}_t+\bepsilon_t$. We measure the dynamic functional connectivity by dynamic correlation, which can be estimated by \gls{bicnet}. Studying dynamic correlation is equivalent to study the dynamic variance, $\text{Var}\left(\by_t\right)=\bLambda\text{Var}\left(\boldsymbol{f}_t\right)\bLambda'$. 

Further, we can see that $\text{Var}\left(\boldsymbol{f}_t\right)$ captures all the network dynamics, and thus we define $\bOmega_t=\text{Var}\left(\boldsymbol{f}_t\right)$ as the \emph{amplitudes} of \glspl{icn}, which captures the dynamics of \gls{icn} activations. Here we adopt the concept of stochastic volatility from economics. Volatility indeed is the variance of a time series, which is also considered a time-dependent random variable \citep{Kim1998,Kastner2019}. For example, in the stock market prices, we can observe that large changes tend to cluster together \citep{Cont2007}, which tells us that the variance of the stock market prices is time-dependent. The stochastic model assumes that the alternation of volatile and tranquil periods is governed by unobservable information flows that are internally generated and never extinguished \citep{HSIEH1991}. This intuition provides a conceptual similarity between financial market fluctuations and brain activities. It is natural to consider that some intrinsic neuronal processes dominate hemodynamic activity measured by \gls{fmri}. These intrinsic neuronal processes are continuously evolving and can become volatile while excited or inhibited by external stimuli or internal demands. Neuroscientists suggest that human brain networks and financial market networks have a certain degree of topological isomorphism \citep{Vertes2011}. Both of these networks have intrinsic properties and respond to external shocks to the system.

Even though \gls{ica} is a very popular method in \gls{icn} studies, it imposes a restriction $\text{Var}\left(\boldsymbol{f}_t\right)=\bI$ to solve the non-identifiability issue \citep{Hyvarinen2000}. Therefore, it only produces spatial maps but not dynamic networks. On the contrary, \gls{bicnet} imposes identifiability restriction on $\bLambda$ which captures the spatial structure of \glspl{icn} but maintain the flexibility of modeling dynamic \gls{icn} amplitudes. There are limited attempts to go beyond \gls{ica} and study the dynamic \glspl{icn}. \cite{Lukemire2020} proposes a statistical framework using Bayesian Gaussian graphical model with a restricted assumption of temporal dependence of \gls{fmri} signals. Thus, there is a need to develop a rigorous statistical model to capture the dynamic nature of connectivity and provide a platform for testing the difference between \glspl{icn} under various conditions, such as across patient groups, across various types of cognitive tasks.

In a recent paper, \cite{Bian2021} discusses methods for change point detection of brain states and a general latent model, which assumes unknown sources and unknown time-varying membership. In contrast, the proposed \gls{bicnet} model focuses on identifying \glspl{icn} with fixed membership and time-varying amplitudes.

Here, we highlight the main difference between the proposed \gls{bicnet} from the \gls{ica}-based methods: (1) \gls{bicnet} can estimate the amplitudes of \glspl{icn}. (2) Existing methods for identifying \glspl{icn} for \gls{fmri} such as \gls{ica} essentially produce co-activation networks (i.e., sets of distributed co-activated brain regions); in contrast, \gls{bicnet} can estimate \underline{both} the co-activation patterns and the strength of connectivity between regions. (3) The sparsity of \glspl{icn} automatically performs feature selection, instead of ad-hoc thresholding. 

The contributions of our proposed \gls{bicnet} model are the following: (1) Its hierarchical structure can simultaneously capture group-level and individual-level variations in \glspl{icn} in a more flexible way. (2) It can quantify the difference of \glspl{icn} amplitudes across different states under a natural statistical inference framework; (3) It identifies multiple \glspl{icn} in a latent subspace, which enables neuroscientists to test for associations between \glspl{icn} with various behavioral measures. (4) It jointly models \gls{rsfmri} and \gls{tfmri} which can produce behaviorally meaningful \glspl{icn}.

The paper is organized as follows. Section 2 introduces the \gls{bicnet} model, including the model specification, MCMC estimation, and inference. Section 3 contains results from extensive simulation studies that investigated the shrinkage behavior of the spike-and-slab prior, the performance of \gls{bicnet} compared with group \gls{ica} algorithm. In Section 4, the proposed model is utilized to analyze the \gls{rsfmri} and \gls{tfmri} data sets from the \gls{hcp}. We conclude with a 
summary and some future extensions of \gls{bicnet} in Section 5.

\section{Methodology: BICNet}

\subsection{Modeling Dynamic Functional Connectivity by Low-Rank Representation}

To begin with, the modeling of dynamic functional connectivity is based on a widely used generalized linear model at voxel level for \gls{fmri} studies, $\widetilde{\boldsymbol{u}}_t = \bB\widetilde{\bx}_t+\widetilde{\by}_t$, where $\widetilde{\boldsymbol{u}}_t$ denotes the original \gls{fmri} signal observed at time $t$, $\widetilde{\bx}_t$ captures the hemodynamic responses, and $\widetilde{\by}_t$ captures functional connectivity in its second or higher order moments. Further, we regress out the activation, $\bB\widetilde{\bx}_t$, since our primary interest is the functional connectivity pattern captured by $\widetilde{\by}_t$. Optionally, we can group the voxels in $\widetilde{\by}_t$ into a \gls{roi} in $\by_t$ using a pre-defined atlas to ease the computation burden. We define $\by_{t,s}^g$ as the $N$-variate \gls{fmri} signals of subject $s$ under experimental condition $g$ with the deterministic mean structure removed. Specifically, $g=0$ refers to the resting state. Therefore, $\by_{t,s}^{0}$ reflects the intrinsic neuronal processes in the lack of external stimuli, and $\by_{t,s}^{g}$, $g\neq 0$, contains information about the alteration of these intrinsic processes evoked by external stimuli.

In the remaining part of this session, we focus on $\by_{t,s}^g$ of an individual $s$ at a given experimental condition $g$, and drop $g$ and $s$ for simplicity. We model the random component $\by_t$ with a conditional Gaussian distribution with time-varying variance $\bSigma_t$, i.e., $\by_t | \bSigma_t \sim \mathcal{N}_N\left(\0,\bSigma_t\right)$, $\forall t\in\{1,\dots,T\}$. Using a latent factor model, we project $\by_t$ into a lower dimensional subspace, $\by_t=\bLambda\boldsymbol{f}_t+\boldsymbol{\epsilon}_t$, where the latent process $\boldsymbol{f}_t|\bOmega_t\sim\mathcal{N}_K(\0,\bOmega_t)$, $K<N$, $\boldsymbol{\epsilon}_{t,s}^g\overset{i.i.d}{\sim}\mathcal{N}_{N}(\0,\bGamma_s^g)$ is the $N\times 1$ isotropic residual with $\bGamma_s^{g}=\text{diag}\left(\sigma_{n,s,g}^{2}\right)$, and $\bLambda$ is the factor loading matrix that captures the time-invariant spatial \gls{icn} structures. It is equivalent to decompose the variance-covariance matrix by $\bSigma_{t} = \bLambda \bOmega_{t} \bLambda^{'} + \bGamma$. Further, we define $\boldsymbol{\lambda}_{k}$ as the $k$th column vector of $\bLambda$, and thus $\bLambda=\left(\boldsymbol{\lambda}_{1},\dots,\boldsymbol{\lambda}_{K}\right)$. Also, we assume the $K$ latent processes are uncorrelated and thus $\bOmega_{t}=\text{diag}\left(\exp(h_{k,t})\right)$.

\begin{figure}[htbp]
\centering
\includegraphics[width=.9\textwidth]{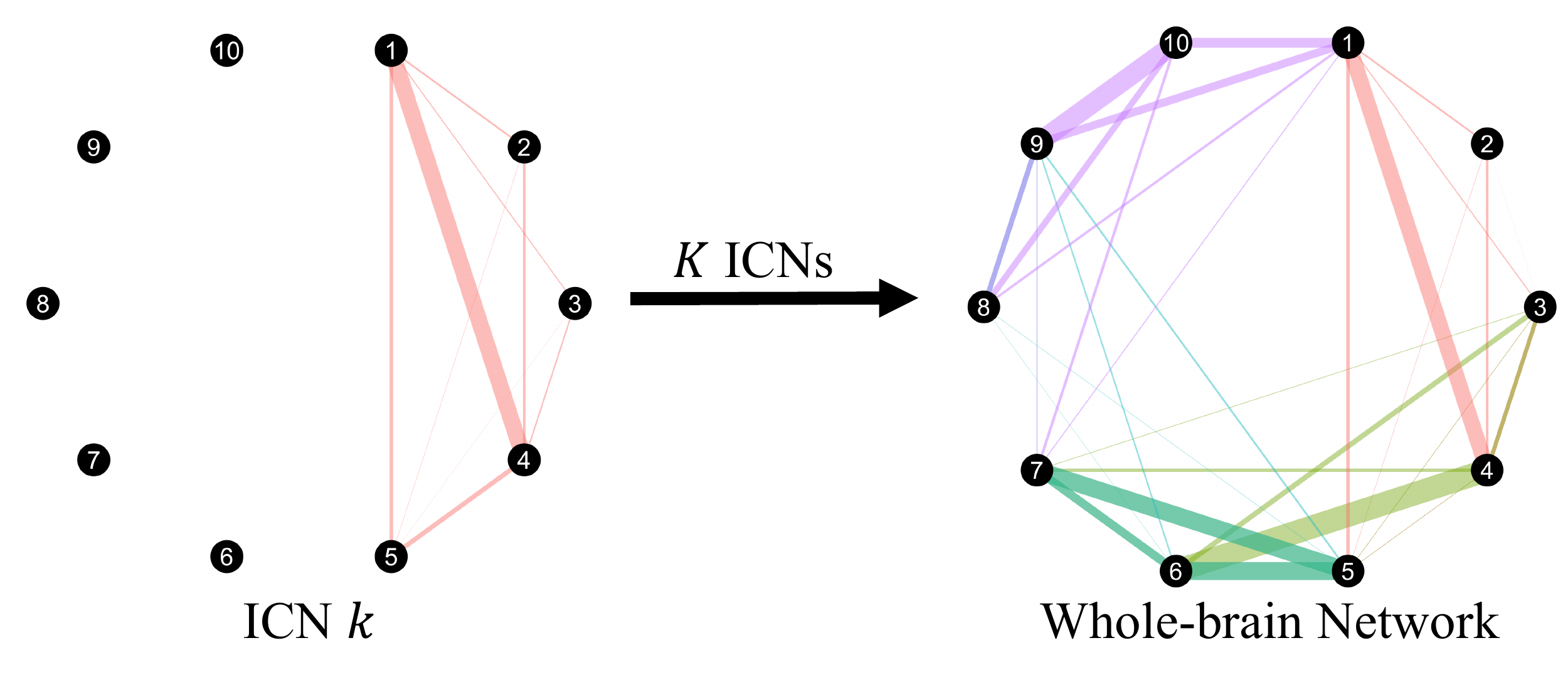}
\caption{Relationship between \glspl{icn} and whole-brain functional connectivity.}
\label{fig:FCNandICN}
\end{figure}

Specifically, the denoised dynamic functional connectivity is defined as $\bLambda\bOmega_{t}\bLambda'$=$\sum_k \exp\left(h_{k,t}\right)\boldsymbol{\lambda}_{k}\boldsymbol{\lambda}_{k}'$. As shown in \cref{fig:FCNandICN}, this denoised whole-brain functional connectivity can be decomposed into $K$ \glspl{icn} with changing dynamics of their corresponding amplitudes. We define \gls{icn} $k$ as $\exp(h_{k,t})\boldsymbol{\lambda}_k\boldsymbol{\lambda}_k^{'}$, where $\exp(h_{k,t})$ is the dynamic amplitude of \gls{icn} $k$, and $\bLambda_{k}$ captures the spatial structure or the co-activation pattern of \gls{icn} $k$, that is, the membership of \gls{icn} $k$ and the contribution of each \gls{roi} that belongs to this \gls{icn}. From another perspective, \gls{icn} $k$ is an undirected graph where $\boldsymbol{\lambda}_k$ defines a set of vertices and their corresponding weights, $\boldsymbol{\lambda}_k\boldsymbol{\lambda}_k^{'}$ defines the edges between vertices, and $\exp(h_{k,t})$ controls the overall weights of edges.

\subsection{Modeling the Structure of ICN by Sparse Hierarchical Model}

In this section, we model both individual \glspl{icn} and the corresponding group-level spatial maps. We add subscript $s$ to indicate a subject $s$ but drop the superscript $g$ since the \gls{icn} structure is invariant across different experimental conditions.

As previously mentioned, the \gls{icn} structure $\boldsymbol{\lambda}_{k,s}$, $\forall k$, consists of membership and the contribution of each member in this \gls{icn}. The zero-nonzero patterns, $\boldsymbol{z}_{k,s}=\left|\boldsymbol{\lambda}_{k,s}\right|>0$ represent the membership of the $k$-th \gls{icn}. A real non-zero value, $\lambda_{n,k,s}\neq 0$, represents the contribution of a \gls{roi} $n$ to an \gls{icn} $k$. While the \gls{icn} memberships can overlap with each other, the dynamics of the $k$-th \gls{icn} are uncorrelated to other \glspl{icn}' dynamics, given that $\bOmega_{t,s}$ is a diagonal matrix. Note here that \glspl{icn} are defined in the latent subspace instead of the whole-brain \gls{roi} space. 

\begin{figure}[htbp]
\centering
\includegraphics[width=.9\textwidth]{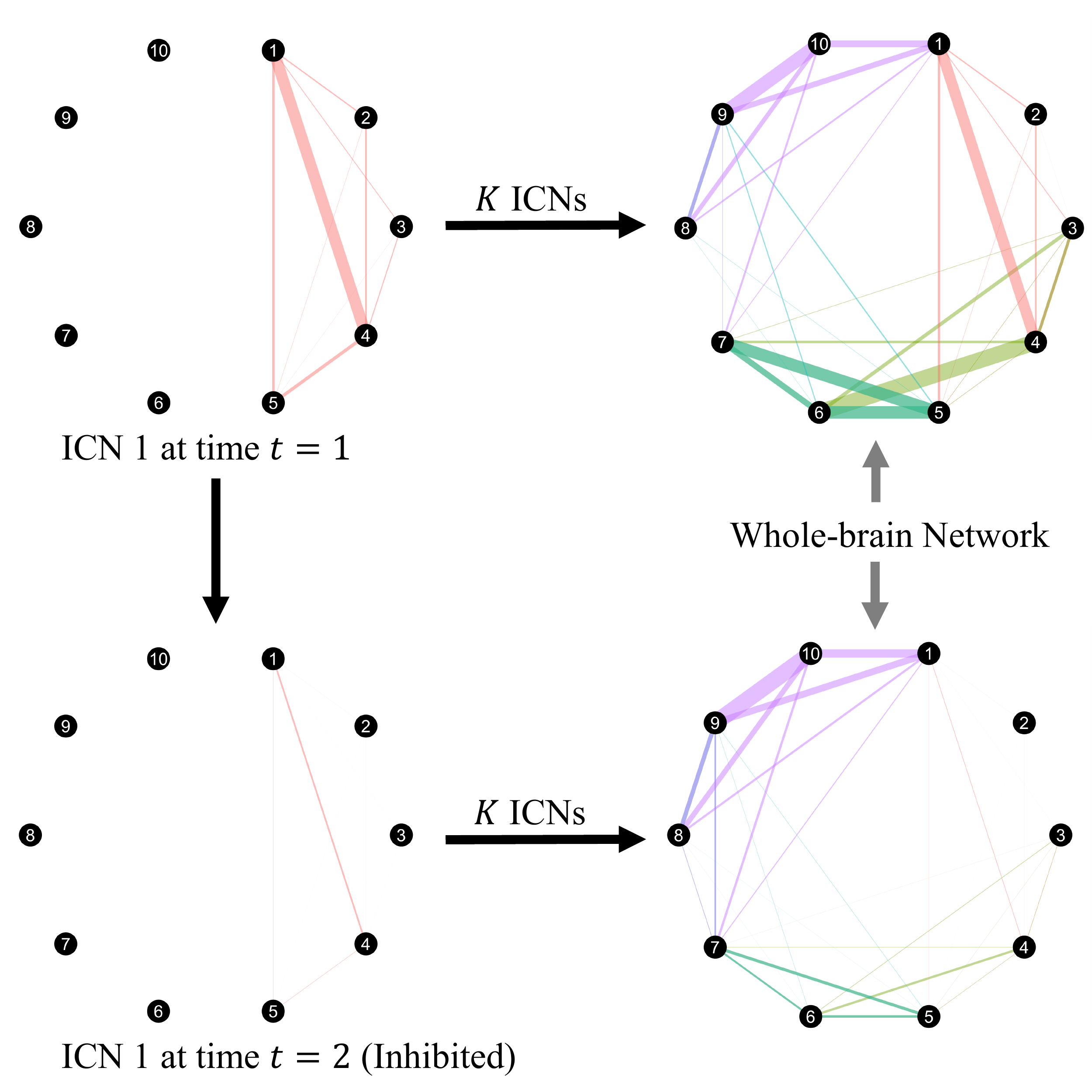}
\caption{Illustration of dynamic \glspl{icn}. An \gls{icn} is an unobservable component of a dynamic functional network. \gls{icn} $k$ is defined as $\exp\left(h_{k,t,s}^g\right)\boldsymbol{\lambda}_{k,s}\boldsymbol{\lambda}_{k,s}'$, where $\exp\left(h_{k,t,s}^g\right)$ is its amplitude at time $t$. The $k$-th column of the factor loading matrix $\bLambda_s$, $\boldsymbol{\lambda}_{k,s}$, represents the structure of the $k$-th \gls{icn}. Specifically, the zero-nonzero patterns, $\boldsymbol{z}_{k,s}=\delta\left(\left|\boldsymbol{\lambda}_{k,s}\right|>0\right)$, represent the membership of the $k$-th \gls{icn}. A real non-zero value, $\lambda_{n,k,s}\neq 0$, represents the contribution of a \gls{roi} $n$ to an \gls{icn} $k$. The change of denoised functional brain network can be decomposed into changes of amplitudes of its \gls{icn} components. For example, from $t=1$ to $t=2$, the structures of \glspl{icn} remain the same while the amplitudes of some \glspl{icn} change, which results in the change of dynamic functional brain networks.}
\label{fig:icnexample}
\end{figure}

The structure of an \gls{icn} is time-invariant and task-invariant. Mathematically, this assumption is necessary to ensure the identifiability of the connectivity parameters. Biologically, this assumption indicates that the structure of every \gls{icn} component is subject-specific and relatively stable across time. However, the \gls{bicnet} model enables dynamic modifications on the amplitudes of \gls{icn} components, which can capture a wide range of global \gls{icn} structures. As shown in the second column of \cref{fig:icnexample}, the structure of \gls{icn} 1 remains the same while the amplitude changes from time $t=1$ to $t=2$. The changes in the amplitudes of the \gls{icn} components can lead to a reconfiguration of the whole-brain functional connectivity networks, as shown in the last column of \cref{fig:icnexample}. Furthermore, this specification is consistent with many studies that suggest that the \gls{icn} structures are relatively stable across conditions. Finally, this model does not constrain the intrinsic connectivity network, $\exp(h_{k,t,s})\boldsymbol{\lambda}_{k,s}\boldsymbol{\lambda}_{k,s}^{'}$, to be constant in time because the state-specific factor variance matrix $\bOmega_{t,s}^{g}$ is allowed to change over time. Therefore, the advantage of \gls{bicnet} is its ability to capture the dynamic nature of brain connectivity.

In the proposed model, the structure of \glspl{icn}, $\bLambda_s$, is sparse in the sense that a specific region is a member of some \glspl{icn}, and an \gls{icn} only includes a subset of regions. To achieve sparsity, we use the standard spike-and-slab prior \citep{George1993,Yu2016a} consisting of a Dirac ``spike'' component that puts a probability mass at zero, and a normal ``slab'' component has its mass spread over a wide range of possible values, as shown in \cref{eq:LambdaInd}. 
\begin{align}
\bLambda_s\text{: } &\lambda_{n,k,s} \sim (1-z_{n,k,s})\delta_0\left(\lambda_{n,k,s}\right) + z_{n,k,s}\mathcal{N}\left(\lambda_{n,k,s}|0,\tau^2\right),\label{eq:LambdaInd}\\
&z_{n,k,s} \overset{i.i.d}{\sim} \text{Bernouli}(\pi_{n,k}), \forall s,\\
\boldsymbol{\Pi}_0\text{: }& \pi_{n,k}\ \ \overset{i.i.d}{\sim}\text{Beta}\left(c\cdot a_k,c \cdot \left(1-a_k\right)\right),\label{eq:LambdaGrp}
\end{align}
\noindent where $z_{n,k,s}=1$ denotes that region $n$ is included in the $k$-th \gls{icn} for subject $s$, while $z_{n,k,s}=0$ indicates otherwise. We assume $z_{n,k,s}\overset{i.i.d}{\sim}\text{Bernouli}(\pi_{n,k})$, $\forall s$, which means that the probability of inclusion of region $n$ in the $k$-th \gls{icn} is denoted to be $\pi_{n,k}$. 

The model currently assumes that $\pi_{n,k}$ is constant across subjects and call it group inclusion probability at $(n,k)$. Here, ``group'' refers to the group of all the subjects. The group inclusion probability matrix, $\boldsymbol{\Pi}_0$, captures the group-level probabilistic spatial maps. As shown in \cref{eq:LambdaGrp}, a beta distribution is utilized to express our uncertainty about this group inclusion probability. The prior expectation of $\pi_{n,k}$ is $\mathbb{E}\left(\pi_{n,k}|a_{n,k}\right)=a_{n,k}$. With this setting, neuroscientists can use biological or experimental information by setting $\mathbf{A}=\left(a_{n,k}\right)$ as well-known \gls{icn} templates, \gls{ica} spatial maps, or \gls{pca} results. Otherwise, a non-informative uniform prior, $\pi_{n,k}\overset{i.i.d}{\sim}\text{Beta}(1,1)$ is used.

The spike-and-slab prior represents the relationship between individual \gls{icn} $\bLambda_{s}$ and the group-level spatial map $\boldsymbol{\Pi}_0$. The \gls{icn} $k$'s structure of subject $s$ has the same mathematical meaning as the \gls{icn} $k$'s structure of subject $l$, $\forall s,l$, since they are all derived from the group-level spatial map $k$. Note that while the probability is common across all subjects, it can flexibly adapt to variations across subjects. The variation in individual \gls{icn} membership is captured by the indicator $\boldsymbol{z}_{k,s}$, and the variation in the contribution of \gls{roi} $n$ to \gls{icn} $k$ is captured by $\lambda_{n,k,s}$.

\subsection{Modeling the Amplitude of ICN by Factor Stochastic Volatility}
\label{sec:modAmp}

We consider a dynamic functional brain network as a mixture of \glspl{icn} that can be either tranquil or volatile. It is natural to assume that when a specific \gls{icn} is activated, either excited or inhibited, its amplitude will first be more volatile than the baseline level and then gradually returns to the baseline level. Specifically, when an \gls{icn} $k$ is excited or inhibited, the amplitude, measured by $\exp\left(h_{k,t,s}^g\right)$, either increases or decreases. We consider both excitation and suppression as activation. Inactivation is defined as no significant change in amplitudes compared to those in the resting state. The dynamics of \gls{icn} amplitudes can be model by the factor stochastic volatility model in \cref{eq:omegaFSV}, which is proposed by \cite{Kastner2017}.

\begin{equation}
\begin{aligned}
f_{k,t,s}^g\ |\ h_{k,t,s}^g &\sim \mathcal{N}\left(0,\exp\left(h_{k,t,s}^g\right)\right)\\
h_{k,t,s}^{g}\ |\ h_{k,t-1,s}^{g},h_{k,t-2,s}^{g},\dots &\sim \mathcal{N}\left( \mu_{k,s}^{g} +\phi_{k,s}^{g}\left(h_{k,t-1,s}^{g}-\mu_{k,s}^{g}\right), \delta_{k,s,g}^{2}\right),
\end{aligned}
\label{eq:omegaFSV}
\end{equation}
\noindent where $k\in\{1,\dots,K\}$, $t\in\{1,\dots,T_g\}$, $s\in\{1,\dots,S\}$, $g\in\{0,\dots,G\}$, and $|\phi|<1$. 

The \gls{ar}(1) process is used to model the dynamics of variance at the logarithm scale, capable of capturing realistic hidden brain states that are nonlinear and continuous. Due to the short length of \gls{fmri} time series, it is common to use \gls{ar}(1) rather than higher-order \glspl{ar} and these are often known to sufficiently capture the temporal structure \citep{Lindquist2008,Gorrostieta2012,Yu2016a}.

\subsection{Modeling the Behavioral Responses by Sparse Linear Regression}

One of the main interests in neuroscience studies is to identify behaviorally meaningful specific \gls{icn} components. Our proposed approach allows associations between subject-specific changes of amplitudes of certain \gls{icn} components and behavioral response under experimental condition $g$ which is denoted by $\bz^g\in\mathbb{R}^S$. These behavioral responses include accuracy, response time, or difficulty level of a specific task. 

As shown in \cref{eq:omegaFSV}, the expected amplitude of latent component $k$ under a specific condition $g$ for a single subject $s$ is captured by $\mu_{k,s}^{g}$. If the true values of $\left\{\mu_{k,s}^g,\forall g\right\}$ is known, then the effect of a task-related stimulus $g$ on a single latent factor $k$ can be quantified by the degree of deviation of $\mu_{k,s}^{g}$, $g\neq 0$, from $\mu_{k,s}^0$. Since the true values are not known, the deviation between the posterior distributions of $\mu_{k,s}^{g}$, $g\neq 0$ and $\mu_{k,s}^0$ is measured using the two-sided two-sample \gls{ks} test statistic, denoted by $\Delta_{ks}^g$, $g\neq 0$. The \gls{ks} statistic is a valid metric because it is symmetric, positive valued (it lies in the unit interval $[0,1]$), and satisfies the triangle inequality. A large value of $\Delta_{ks}$ indicates \gls{icn} $k$ of subject $s$ is activated. To further distinguish inhibition and excitation under activation, we calculate the sign of the difference between the posterior means, $\widetilde{\Delta}_{ks}=\text{sgn}\left(\overline{\mu}_{ks}^1-\overline{\mu}_{ks}^0\right)$, with $\widetilde{\boldsymbol{\Delta}}=\left(\widetilde{\Delta}_{ks}\right)$. Thus, a small value of $\Delta_{ks}$ indicates no activation, whereas a large value of $\Delta_{ks}$ with $\widetilde{\Delta}_{ks}=-1$ indicates inhibition, and a large value of $\Delta_{ks}$ with $\widetilde{\Delta}_{ks}=1$ indicates excitation. 

A regression model is employed to investigate the association between behavioral measures, and the changes of amplitudes of the \glspl{icn} during task performance relative to the resting state, measured by $\boldsymbol{\Delta}^g=\left(\Delta_{k,s}^g\right)$: 
\begin{equation}
\bz^g | \boldsymbol{\Delta}^g, \bbeta^g, \sigma^2_g \sim \mathcal{N}_s\left(\boldsymbol{\Delta}^g\bbeta^g,\sigma^2_g\boldsymbol{I}_S\right),
\end{equation}
where the regression coefficient $\bbeta^g\in\mathbb{R}^K$ measures the contribution of each \glspl{icn}`' change of amplitudes to the behavioral response, $\sigma^2_g$ is the variance of the behavioral measure. The variance parameter $\sigma^2_g$ follows an inverse Gamma distribution, $\text{IG}(\alpha_1,\alpha_2)$. For $\bbeta$, we assume that only a few of \glspl{icn} have contribution to $\boldsymbol{z}^g$, thus we use a spike-and-slab prior on $\bbeta$. 
\begin{align}
&\beta_k | \pi_k,\sigma^2_g,\tau^2 \sim (1-\pi_k)\delta_0\left(\beta_k\right) + \pi_k\mathcal{N}\left(\beta_k|0,\sigma^2_g\tau^2\right)\\
&\pi_k | \theta \sim \text{Bernouli}(\theta),\ k=1,\dots,K
\end{align}
For a behavioral response $\bz^g$, if the 95\% credible interval of $\beta_{k}^g$'s posterior distribution does not contain zero, we say that there is an association between the $k$-th \gls{icn} component and this behavioral response at the group level. From this model, one can identify the \glspl{icn} whose differential effect (task vs. rest) was significantly associated with accuracy, response time, story difficulty level, and math difficulty level, respectively. 
\subsection{Prior Distributions}

For the univariate \gls{sv} processes, the prior distributions proposed by \cite{Kastner2017} 
will be used, i.e., $\mu_{k,s}^g \overset{i.i.d}{\sim}\mathcal{N}(b_{\mu},B_{\mu})$, $\frac{\phi^{g}_{k,s}+1}{2} \overset{i.i.d}{\sim} \text{Beta}(a_{\phi},b_{\phi})$, and $\delta^{2}_{k,s,g} \overset{i.i.d}{\sim} \text{Gamma}\left(\frac{1}{2},\frac{1}{2B_{\delta}}\right)$.
Here, the Gamma distribution for $\delta^{2}_{k,s,g}$ is used instead of the conditionally conjugate inverse Gamma prior. Though the conjugate inverse Gamma prior is more commonly used because of computational efficiency, it bounds the variance away from zero a priori, leading to over-fitting in the state-space model \citep{Simpson2017,Klein2016,Kastner2017}. The Gamma distribution for variance is equivalent to a truncated normal distribution for the standard deviation. There is some evidence suggesting that a Gaussian distribution \citep{Hlinka2011} can reasonably approximate the \gls{rsfmri} process. Therefore this choice prefers the simple normal distribution to the stochastic volatility process and is data-adaptive.

Conversely, the conjugate inverse Gamma is used for the regional-specific variance of the observed signals, $\sigma^2_{n,s,g}\overset{i.i.d}{\sim}\text{IG}(c_{\sigma},d_{\sigma})$, where $c_{\sigma}$ is the shape parameter and $d_{\sigma}$ is the rate parameter. Since \gls{fmri} data tends to have relatively small activation fluctuation compared with its noisy background \citep{Welvaert2013}, 
it is expected that the regional-specific white noise to account for a large part of the variation. 
Hence, the factor stochastic volatility process has a relatively small variation but contains more information about the hidden brain states.
For behavioral responses, $\theta$, which is the probability of having nonzero regression 
coefficients, is modeled as $\text{Beta}(a,b)$, and $\tau^2 | S^2 \sim \text{IG}\left(1/2, S^2/2\right)$.

\subsection{Estimation via MCMC}

\begin{figure}
\centering
\includegraphics[width=.9\textwidth]{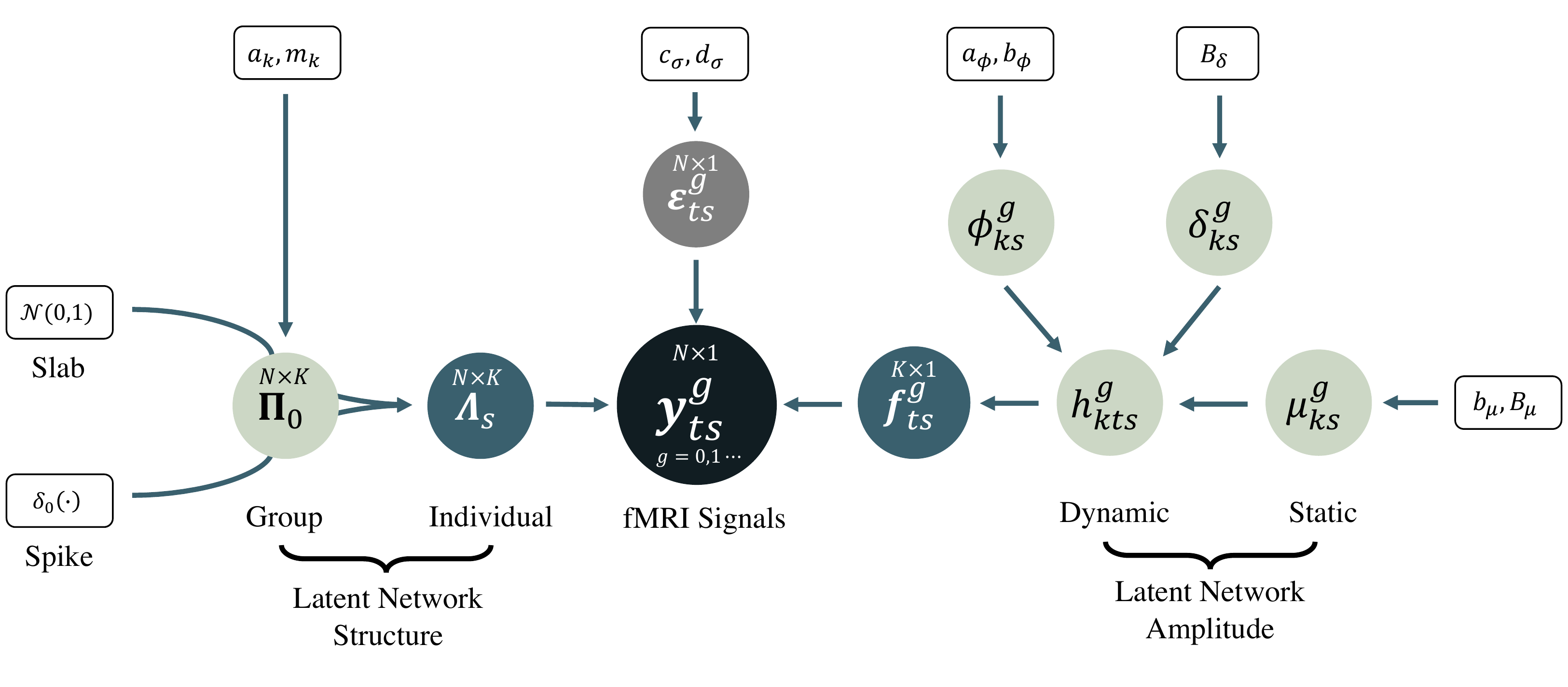}
\caption{The directed acyclic graph of our proposed model. The black round node is the filtered fMRI signal $\by^g_{ts}$. The dark and light green nodes are hidden variables. The white rectangular nodes are hyperparameters of the prior distributions.}
\label{fig:singleDAG}
\end{figure}

A Gibbs sampler augmented by some Metropolis-Hasting steps is constructed for factor stochastic volatility for our proposed model. The \gls{dag} is illustrated in \cref{fig:singleDAG}. The sampling process can be parallelized within each MCMC iterations because many conditional posterior distributions are independent of others. The factor loading matrix signs are not identifiable, and we solve this issue by post-processing after MCMC sampling. The sampling process is summarized as follows. 

\begin{enumerate}
  \item We first update the regional-specific variances of the observed signals. Conditional on the latent factor structure, the observed signals follow a normal distribution, $\by_{t,s}^{g}|\bLambda_s,\boldsymbol{f}_{t,s}^{g}\sim\mathcal{N}\left(\bLambda_s\boldsymbol{f}_{t,s}^{g},\bGamma_s^{g}\right)$. Given the inverse Gamma prior, the conjugate conditional posterior distribution is $\sigma^2_{n,s,g}\ |\ \by_{n,s}^{g}, \bF_s^{g}, \boldsymbol{\lambda}_{n,s}, c_{\sigma}, d_{\sigma}$. The sampling can be parallelized across $g$ and $n$ since they are conditionally independent.
  \item Sample stochastic volatilities of the latent factors, $\{h_{k,t,s}^{g}\}$, and the corresponding parameters $\{\mu_{k,s}^{g}, \phi_{k,s}^{g}, \delta_{k,s,g}\}$ separately for every $k$ and $g$ using the 2-step MH sampler with centered parameterization and interweaved by a non-centered parameterization \citep{Kastner2014}.
  \item For the sparse factor loadings, we first sample the sparse indicator $z_{n,k,s}=\delta(\lambda_{n,k,s}\neq 0)$ based on its conditional posterior odds. If $z_{n,k,s}=1$, then sample $\lambda_{n,k,s}$ from its conjugate conditional posterior distribution, $\lambda_{n,k,s}\ |\ \{\by_{n,s}^{g}\}, \boldsymbol{\lambda}_{n,-k,s}, \{\boldsymbol{f}_{k,s}^{g}\}, \sigma^2_{n,s,g}, \tau^2$; otherwise, set $\lambda_{n,k,s}=0$. 
  \item Sample latent factor $\boldsymbol{f}^{g}_{t,s}|\boldsymbol{h}_{t,s}^{g},\by_{t,s}^{g},\bLambda_s,\sigma^2_{n,s,g}$ using a normal Bayesian regression update.
  \item Sample inclusion probability, or the group \glspl{icn}, using conditional posterior distribution: $\pi_{n,k}\ |\ \{z_{n,k,s},\forall s\}, a_k, c \sim\text{Beta}\left(c\cdot a_k+\sum_s z_{n,k,s}, c\cdot(1-a_k)+S-\sum_s z_{n,k,s}\right)$.
\end{enumerate}

\noindent For the sparse linear regression of behavioral measures, we construct a Gibbs sampler as follows.

\begin{enumerate}
  \item We first update the probability of having nonzero regression coefficients with $$\theta|\boldsymbol{\pi}\sim\text{Beta}\left(a+\sum_k\pi_k,b+\sum_k(1-\pi_k)\right).$$
  \item Then we sample $\tau^2|\bbeta,\boldsymbol{\pi}\sim\text{IG}\left(\frac{1}{2}+\frac{1}{2}\sum_k \pi_k,\frac{1}{2}S^2+\frac{1}{2\sigma^2}\bbeta'\bbeta\right)$.
  \item For the variance of behavioral measure $\bz$, we have $$\sigma^2|\bz,\bbeta\sim\text{IG}\left(\alpha_1+S/2,\alpha_2+\frac{1}{2}(\bz-\boldsymbol{\Delta}\bbeta)'(\bz-\boldsymbol{\Delta}\bbeta)\right).$$
  \item For the regression coefficient $\bbeta$, which measures the contribution of \glspl{icn}, we have $\bbeta|\bz,\sigma^2,\boldsymbol{\pi},\tau^2\sim\mathcal{N}\left(\boldsymbol{\Sigma^{\text{post}}}\boldsymbol{\Delta}'\bz\frac{1}{\sigma^2},\Sigma^{\text{post}}\right)$, where $\Sigma^{\text{post}}=(\frac{1}{\sigma^2}\boldsymbol{\Delta}'\boldsymbol{\Delta}+\frac{1}{\sigma^2\tau^2}\mathbf{I}_K)^{-1}$.
  \item Define $\zeta=\boldsymbol{\Delta}_{k\cdot}'\boldsymbol{\Delta}_{k\cdot}+\tau^{-2}$. For $\pi_k$, which indicates if $\beta_k=0$, we have 
  $$ \text{P}(\pi_k=1|\cdot) = 
  (1-\theta)/\left((\sigma^2\tau^2)^{-1/2}\exp\left(\frac{(\bz'\boldsymbol{\Delta}_{k\cdot})^2\sigma^{-2}}{2\zeta}\right)(\sigma^{-2}\zeta)^{-1/2}\theta+(1-\theta)\right).
  $$
\end{enumerate}

\subsection{Inference}
\label{sec:infer}
\Gls{bicnet} can identify individual \glspl{icn} and group-level spatial maps. However, the group-level probability spatial maps, $\boldsymbol{\Pi}_0$, are non-sparse inclusion probability matrices, while the individual \glspl{icn} are sparse matrices. We threshold $\boldsymbol{\Pi}_0$ at a selected probability value to aid the group-level interpretation. Similarly, in the following sections, we will estimate group \gls{ica} at the \gls{roi}-level using the \gls{gift} algorithm. For the \gls{ica} spatial maps, we calculate the z-scores and threshold them at a selected value.

Identify \glspl{icn} and spatial maps that are related to language processing is equivalent to investigate the association between behavioral measures and the extent of activation (either suppression or excitation) of the \glspl{icn} during task performance relative to the resting state. We fit a linear regression model with behavioral measures as the dependent variables and the $K\times S$ matrix $\boldsymbol{\Delta}=\left(\Delta_{ks}\right)$ as regressors. From the model, one can identify the \glspl{icn} whose differential effect (task vs. rest) was significantly associated with accuracy, response time, story difficulty level, and math difficulty level.

\section{Simulation}

\subsection{Shrinkage Effect on ICNs with Different Underlying Sparsity}
\begin{figure}
\centering
\subfloat[True $\bLambda_s$]{\includegraphics[width=.45\textwidth]{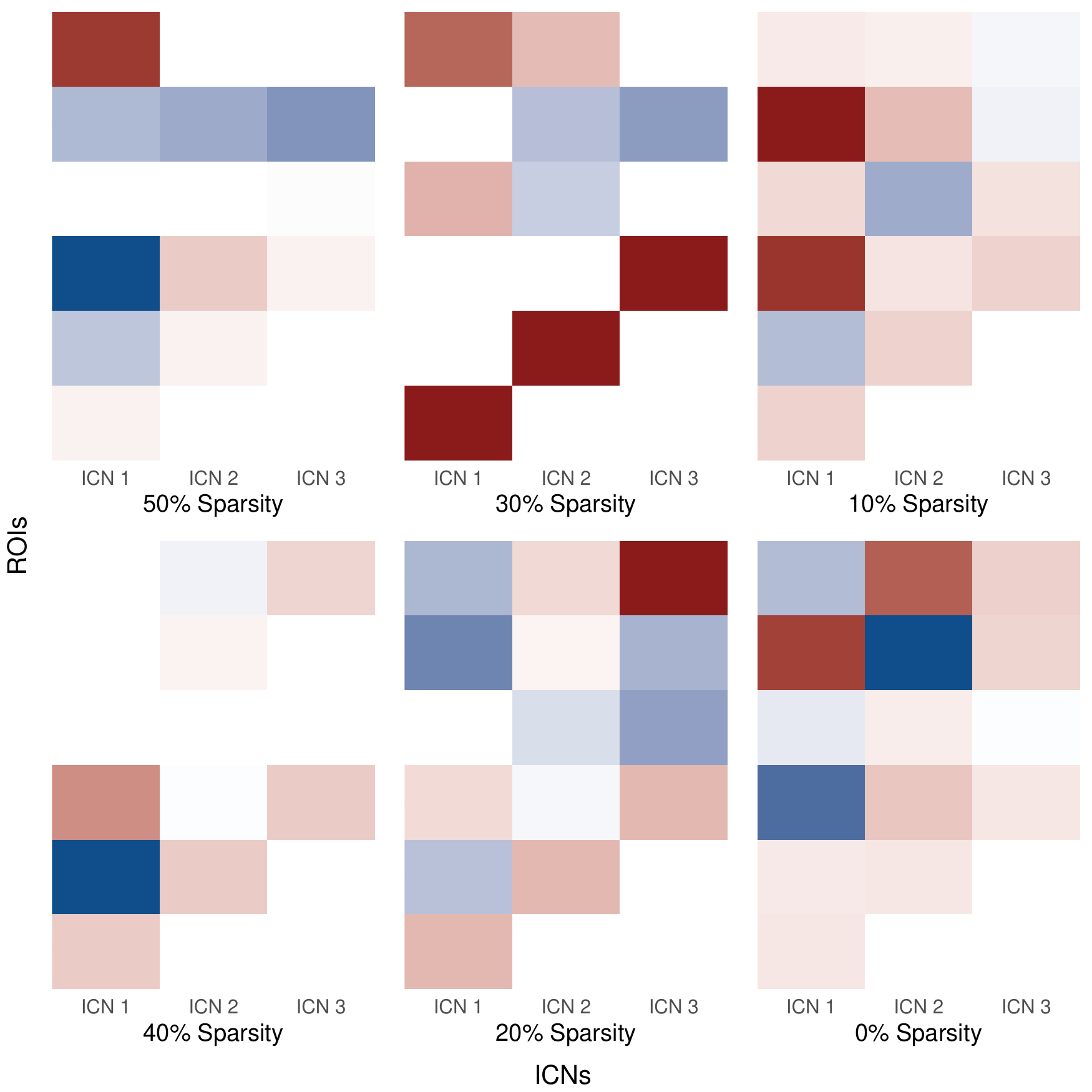}}
\subfloat[\gls{bicnet} $\widehat{\bLambda}_s$]{\includegraphics[width=.45\textwidth]{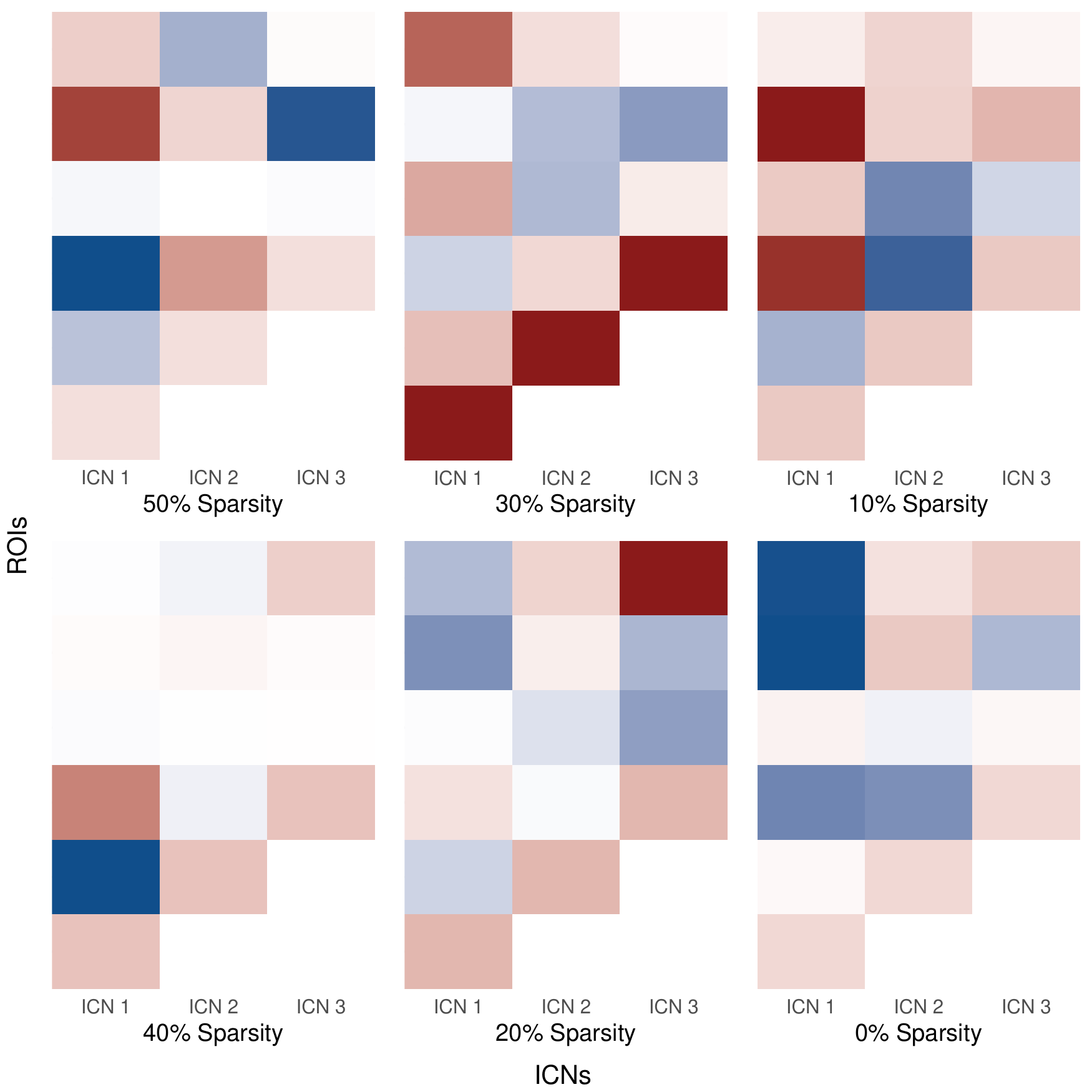}}
\caption{(a) True $\bLambda_s$ for six subjects with different level of sparsity for each subject. (b) \glspl{icn} estimated by \gls{gift}. (b) \glspl{icn} estimated by \gls{bicnet}. The $\widehat{\bLambda}_s$ is the median of 10000 MCMC samples generated from a single-subject \gls{bicnet} model with the first 10000 samples discarded as burn-ins.}
\label{fig:SimMaeSparsepct}
\end{figure}

In the first experiment, the test is on inferring \gls{icn} structures under different sparsity levels. In other words, this experiment investigates the shrinkage effect of the spike-and-slab prior and how well it adapts to different levels of underlying sparsity on $\bLambda_s$.
We simulate a small-scale brain network of $N=6$ \glspl{roi} with $K=3$ latent \glspl{icn} with different levels of sparsity, as illustrated in \cref{fig:SimMaeSparsepct}(a). To control the underlying sparsity, each subject has different \gls{icn} membership and the proportion of nonzero loadings in the factor loading matrix varies from 50\% to 100\%, with an increment of 0.1. For each \gls{icn}, the expectation of the logarithmic amplitude is $\mu_{k,s}^{1}=2-k$, $k\in\{1,2,3\}$. We fix $\phi=0.9$ and $\delta=0.5$ for all subjects and experimental conditions. Each experimental condition has $T_g=1000$ temporal observations and results in 2000 temporal observations in total. 
The accuracy of estimated \gls{icn} structure is measured by \gls{mae}, defined as $\text{MAE}_s=\sum_n\sum_k|\Lambda_{n,k,s}-\hat{\Lambda}_{n,k,s}|/(NK)$. We also measure the accuracy of reconstructed time series by \gls{rmse}, defined as $\text{RMSE}_s = \sqrt{\sum_t\left(\widehat{\bLambda}_s \widehat{\boldsymbol{f}}_{t,s} - \bLambda_s \boldsymbol{f}_{t,s}\right)^2/T}$.

We use the single-subject \gls{bicnet} model without the layer of group inclusion probability because the subjects are designed to have very different \gls{icn} structures. The estimator is the median of 10000 MCMC samples, with the first 10000 samples discarded as burn-in. It is obvious in \cref{fig:SimMaeSparsepct} that the estimated \gls{icn} structures recover most of the zero-nonzero patterns in the true \gls{icn} structures. We generated 50 simulated datasets and applied the single-subject \gls{bicnet} respectively. As shown in \cref{tab:SimAccSparsePct}, there is no significant trend in the \gls{rmse} and \gls{mae} as the non-sparsity increases, which indicates the spike-and-slab prior can be well adapted to different individual sparsity.

\begin{table}[]
\centering
\begin{tabular}{@{}lll@{}}
\toprule
Pct. of Non-sparsity & RMSE              & MAE               \\ \midrule
50\%                 & 0.4489$\pm$0.0146 & 0.0226$\pm$0.0082 \\
60\%                 & 0.4408$\pm$0.0220 & 0.0441$\pm$0.1167 \\
70\%                 & 0.4596$\pm$0.0456 & 0.0610$\pm$0.2133 \\
80\%                 & 0.4524$\pm$0.0159 & 0.0317$\pm$0.014  \\
100\%                & 0.4574$\pm$0.0175 & 0.0468$\pm$0.1079 \\ \bottomrule
\end{tabular}
\caption{Accuracy of \gls{bicnet} estimation under different percentage of underlying true sparsity. The estimator is the median of 10000 MCMC samples with the first 10000 samples discarded as burn-ins. We use a single-subject \gls{bicnet} model without the layer of group inclusion probability. The mean and standard deviation are calculated with the \gls{rmse} and \gls{mae} from the 50 repetitions. In each repetition, we simulate a new dataset under the same setting and apply the single-subject \gls{bicnet} model.}
\label{tab:SimAccSparsePct}
\end{table}

\subsection{BICNET vs. Group ICA}
In the second experiment, we simulate a large-scale brain network of $N=90$ \glspl{roi} with $K=18$ latent \glspl{icn}, $T=200$ observations, and $S=20$ subjects. All subjects have the same \gls{icn} membership and come from a homogeneous group categorized by a group inclusion probability $\boldsymbol{\Pi}_0$. For \gls{bicnet} estimation, we use the multi-subject model without any constraints on $\left\{{\bLambda_s},\forall s\right\}$, the \gls{icn} structure, with 10000 MCMC samples after discarding the first 20000 samples as burn-ins. We run 5 MCMC chains with random initial values. We compare the estimation performance of individual \glspl{icn} with \gls{mae} defined above and \gls{rmse} defined as $\text{RMSE}_s = \sqrt{\sum_t\left(\widehat{\bLambda}_s \widehat{\boldsymbol{f}}_{:t,s} - \bLambda_s \boldsymbol{f}_{:t,s}\right)^2/T}$. Also, we use \gls{mae} to compare the performance of group inclusion probability estimation.

\begin{table}[]
\centering
\begin{tabular}{@{}llll@{}}
\toprule
             & \multicolumn{2}{l}{$\bLambda_s$} & $\boldsymbol{\Pi}_0$\\ \midrule
             & \gls{mae}     & \gls{rmse}    & \gls{mae} \\
\gls{bicnet} & $0.47\pm0.04$ & $1.04\pm0.19$ & 0.57\\
\gls{gift}   & $0.78\pm0.02$ & $3.90\pm0.39$ & 0.76 \\ \bottomrule
\end{tabular}
\caption{Performance of \gls{bicnet} and \gls{gift} on a large-scale simulated dataset. We use a multi-subject \gls{bicnet} model without any constraints on $\left\{{\bLambda_s},\forall s\right\}$. We use the median of the 10000 MCMC samples as estimation after discarding the first 20000 samples as burn-in.}
\label{tab:SimMaeLarge}
\end{table}

\begin{figure}
\centering
\subfloat[Ture $\bLambda_s$]{\includegraphics[width=.2\textwidth]{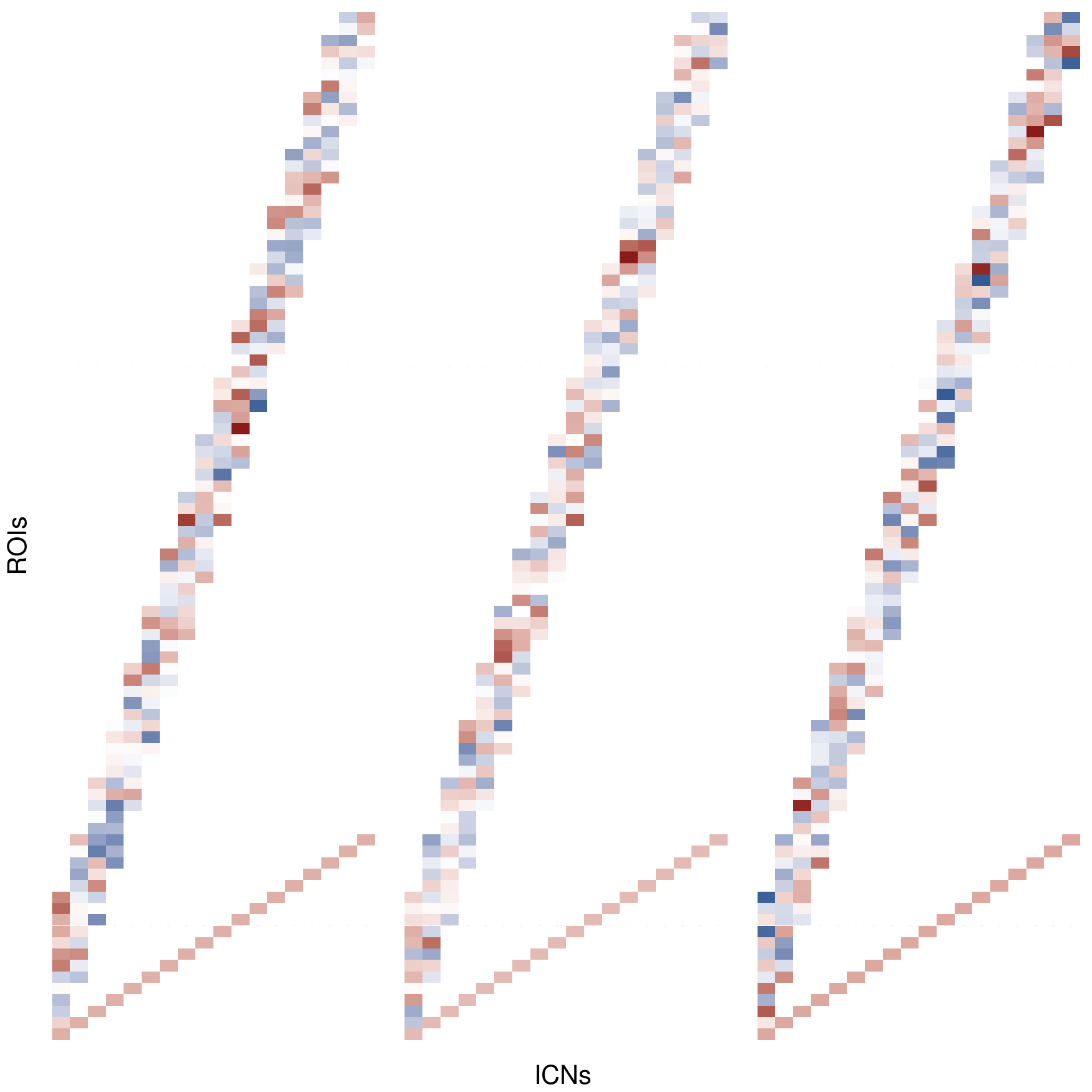}}
\subfloat[\gls{gift}]{\includegraphics[width=.2\textwidth]{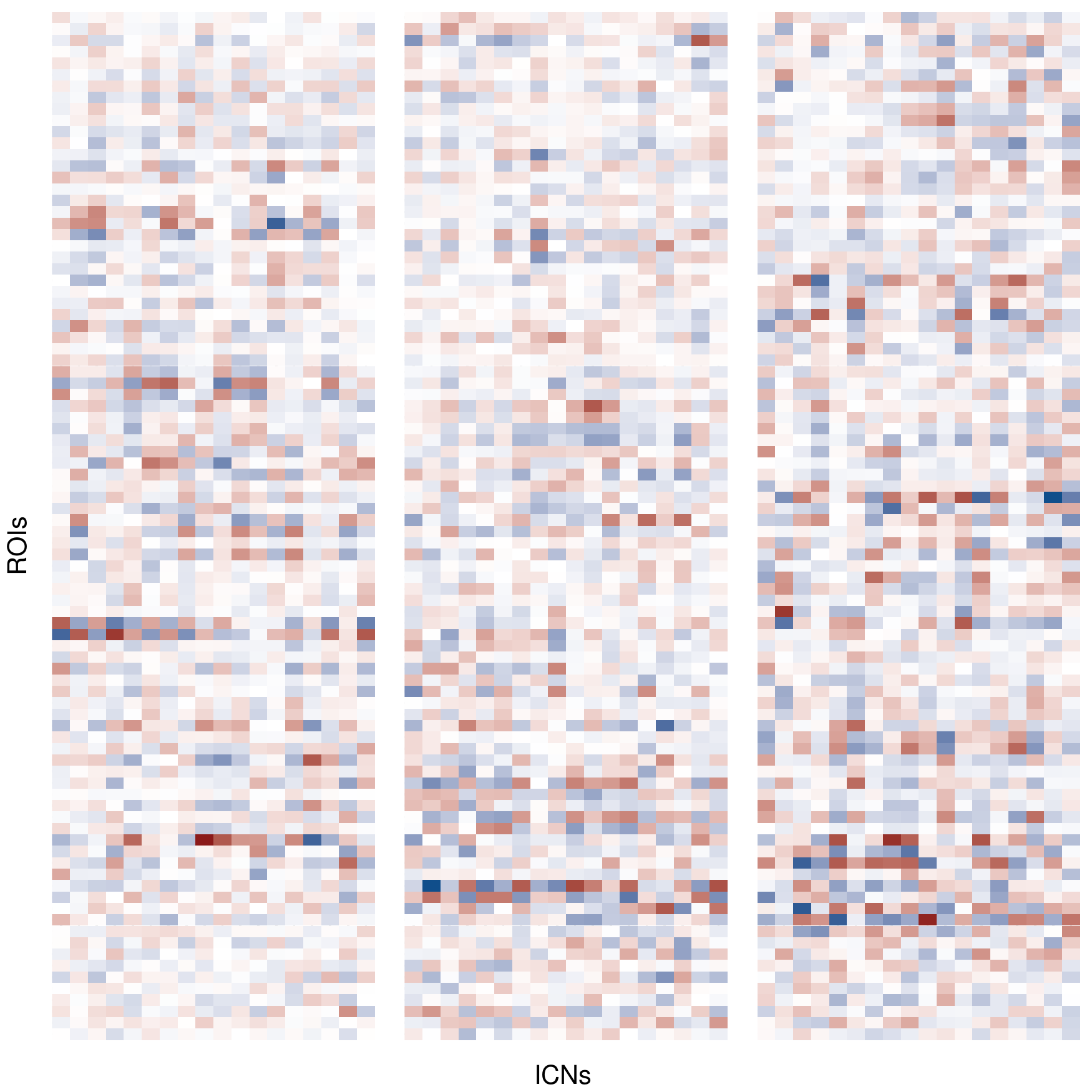}}
\subfloat[\gls{bicnet} Non-Permuted]{\includegraphics[width=.2\textwidth]{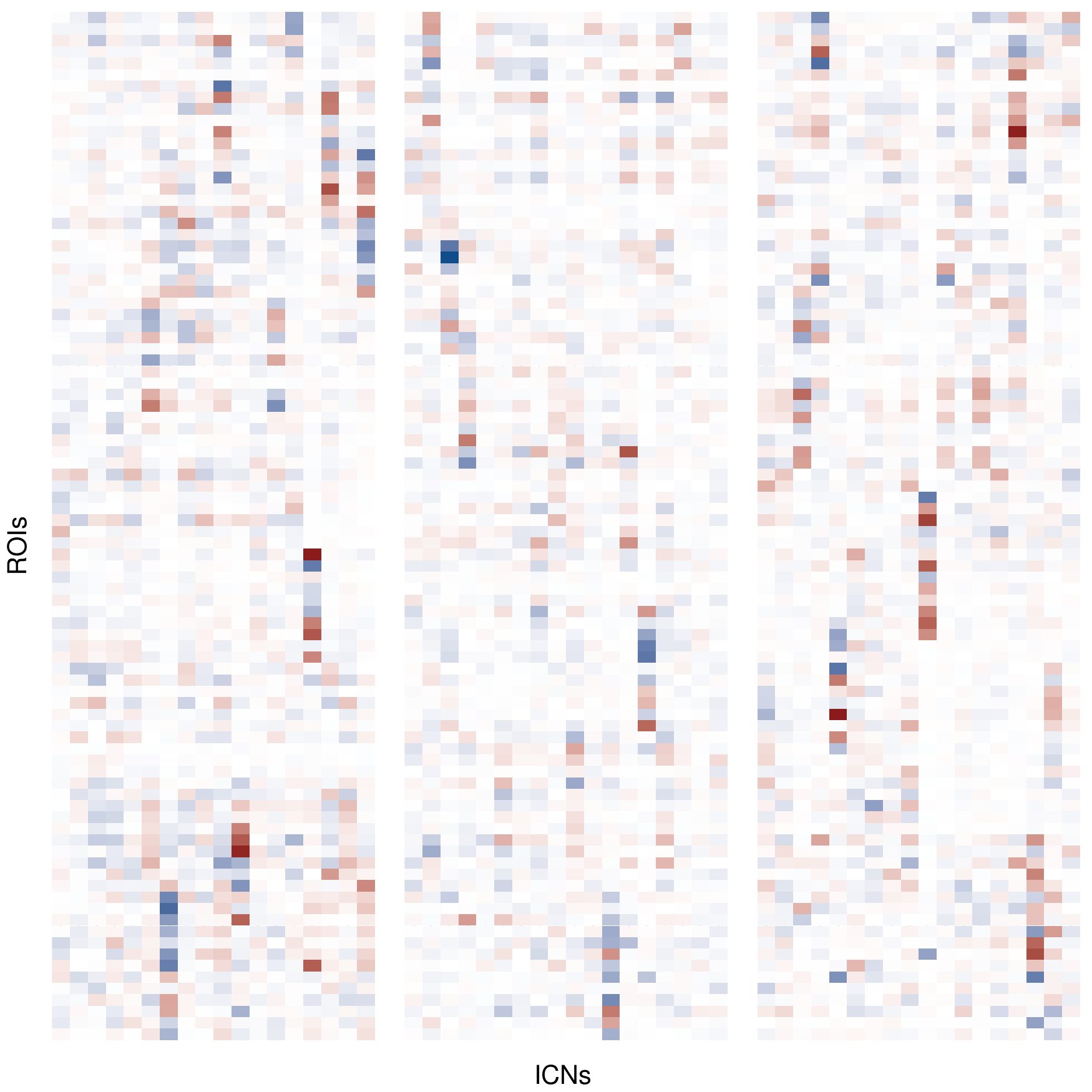}}
\subfloat[\gls{bicnet} Permuted]{\includegraphics[width=.2\textwidth]{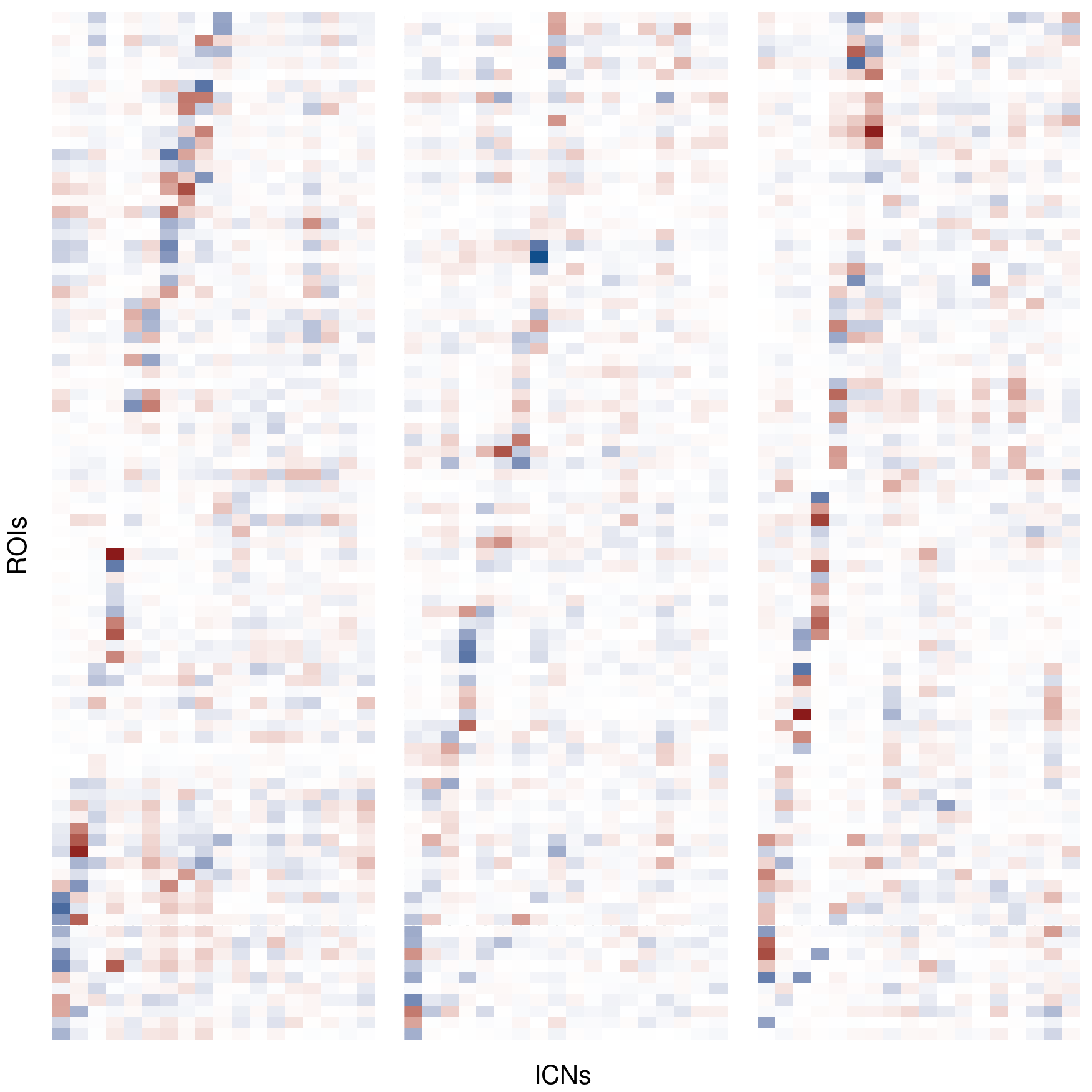}}
\caption{Estimation of the simulated large-scale brain network of $N=90$ \glspl{roi} with $K=18$ latent \glspl{icn}, $T=200$ observations, and $S=20$ subjects. Only three subjects are presented for figures (a)-(d). (a) True $\bLambda_s$ for three subjects. (b) \glspl{icn} estimated by \gls{gift}. (c) \glspl{icn} estimated by \gls{bicnet}. This estimation suffers from the column switching issue. (d) To solve the column switching issue, we permute \glspl{icn} estimated by \gls{bicnet}.}
\label{fig:SimMaeBig}
\end{figure}

As shown in \cref{tab:SimMaeLarge}, \gls{bicnet} has a higher accuracy in estimating the \gls{icn} structures and the denoised whole-brain networks. \Cref{fig:SimMaeBig} shows that the $\widehat{\bLambda}_s$ estimated by \gls{bicnet} and \gls{gift}. The \gls{bicnet} estimation captures more sparse structure than \gls{gift}. 

\section{Modeling the ICN in fMRI Data}

\subsection{Data}
The \acrfull{hcp} \citep{VanEssen2012} collected multi-modal imaging data from 1200 subjects. The imaging modalities include \gls{rsfmri} and \gls{tfmri} under different experimental settings. This project provides a massive, comprehensive database that can shed light on the human brain's anatomical and functional structure. Specifically, neuroscientists are interested in partitioning the brain into neurobiologically and functionally meaningful areas and relating them to our cognitive behaviors \citep{Glasser2016}. Neuroscientists believe that studying functional brain networks can underpin individual differences, neurological disorders, and aging.

We analyzed the \gls{rsfmri} and \gls{tfmri} collected from a language task. During the scanning of \gls{rsfmri}, participants kept their eyes open and fixed on a bright cross with a dark background. In the language task, participants were asked to finish four blocks of a story task and four blocks of a math task alternatively. In the story task, the participants listened to brief auditory stories and answered this story's main idea. In the math task, the participants listened to arithmetic questions, e.g., simple addition and subtraction. For both tasks, participants need to answer single-choice auditory questions by pressing a button.

The data were acquired with 3T Siemens Skyra with TR = 720 ms, TE = 33.1 ms, flip angle = $52^{\circ}$, BW =2290 Hz/Px, in-plane FOV = 208 x 180 mm, 72 slices, 2.0 mm isotropic voxels. The \gls{rsfmri} has 1200 frames per run, approximately 15 minutes per run. Each run of the language \gls{tfmri} has 316 frames, approximately 4 minutes. We use the minimally preprocessed data of 200 subjects from the \gls{hcp} database. The \gls{hcp} minimal preprocessing pipeline \citep{Glasser2013} includes a correction for B0 distortion, realignment to correct for motion, registration to the participant's structural scan, normalization to the 4D means, brain masking, and nonlinear warping to MNI space. For each voxel, we remove the hemodynamic response and normalize the remained signal in a voxel-wise manner. Later, we partition the voxels into 90 \glspl{roi} using the \gls{aal}, and use the sample mean to summarize \gls{fmri} signals within each \gls{roi}. As listed in \cref{tab:behav}, language \gls{tfmri} comes with behavioral measures, including accuracy, response time, story tasks' difficulty levels, and math tasks' difficulty levels. 

\begin{table}[htbp]
\centering
\begin{tabular}{@{}rrrrr@{}}
\toprule
Subject & Accuracy & Resp. Time & Story Diff. Lv. & Math Diff. Lv. \\ \midrule
1 & 1.102586  & -1.234231 & 9.071  & 1.964 \\
2 & -1.039474 & -0.301118 & 9.607  & 2.893 \\
3 & 1.509315  & -0.020224 & 12.446 & 3.107 \\
4 & -0.623765 & -0.480615 & 11.393 & 1.571 \\
5 & 0.424749  & 1.196522  & 9.375  & 1.964 \\ \bottomrule
\end{tabular}
\caption{Individual behavioral measures of language tasks. The measures are accuracy, response time, story difficulty level and math difficulty.}
\label{tab:behav}
\end{table}

\subsection{Model Settings}

\begin{figure}
\centering
\subfloat[$\mu_{ks}^g\sim\mathcal{N}(b_{\mu},B_{\mu})$]{\includegraphics[width=.3\textwidth]{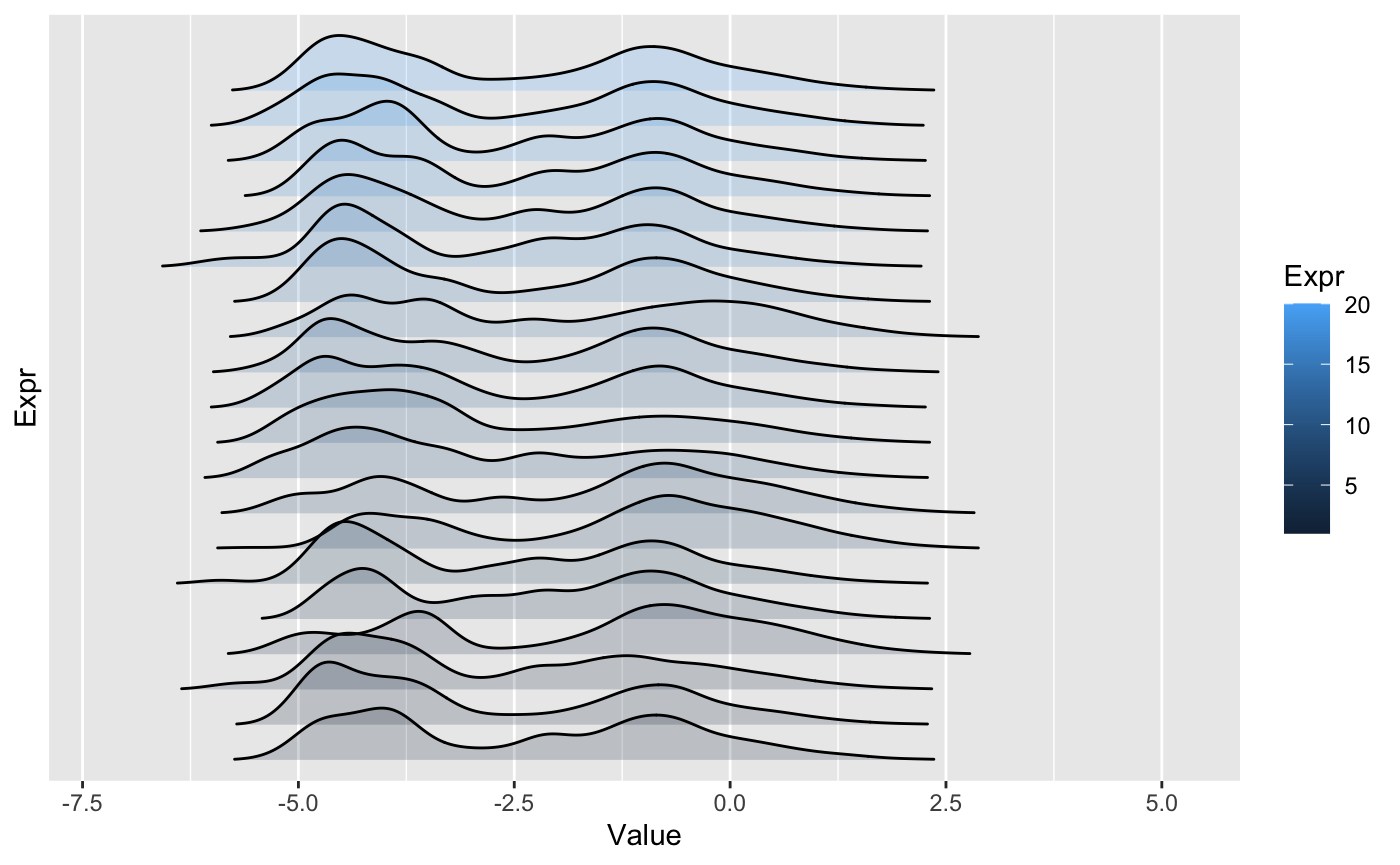}}
\subfloat[$(\phi_{ks}^g+1)/2\sim\text{Beta}(a_{\phi},b_{\phi})$]{\includegraphics[width=.3\textwidth]{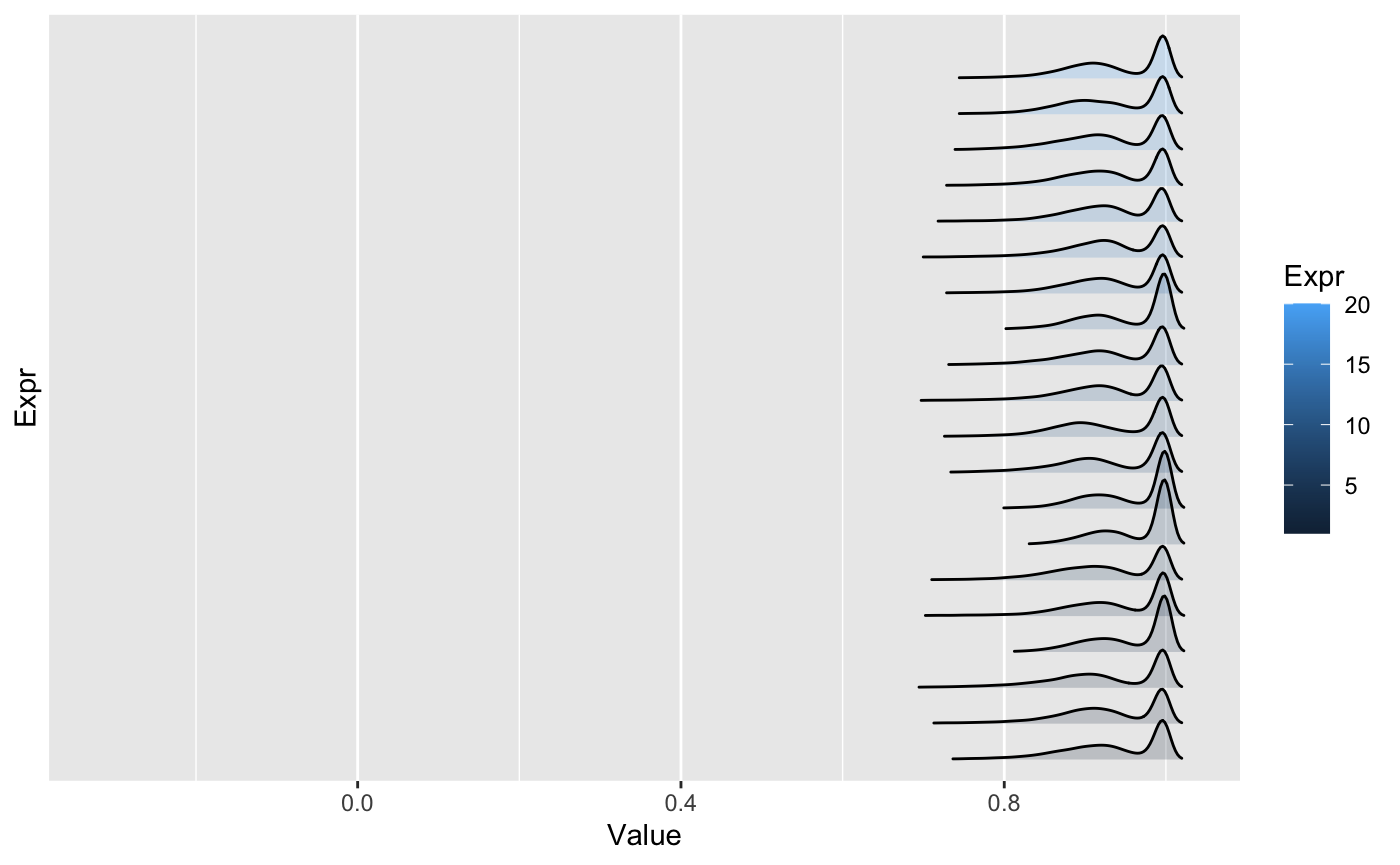}}
\subfloat[$\delta_{ksg}^2\sim\text{Gamma}(0.5,0.5B_{\sigma})$]{\includegraphics[width=.3\textwidth]{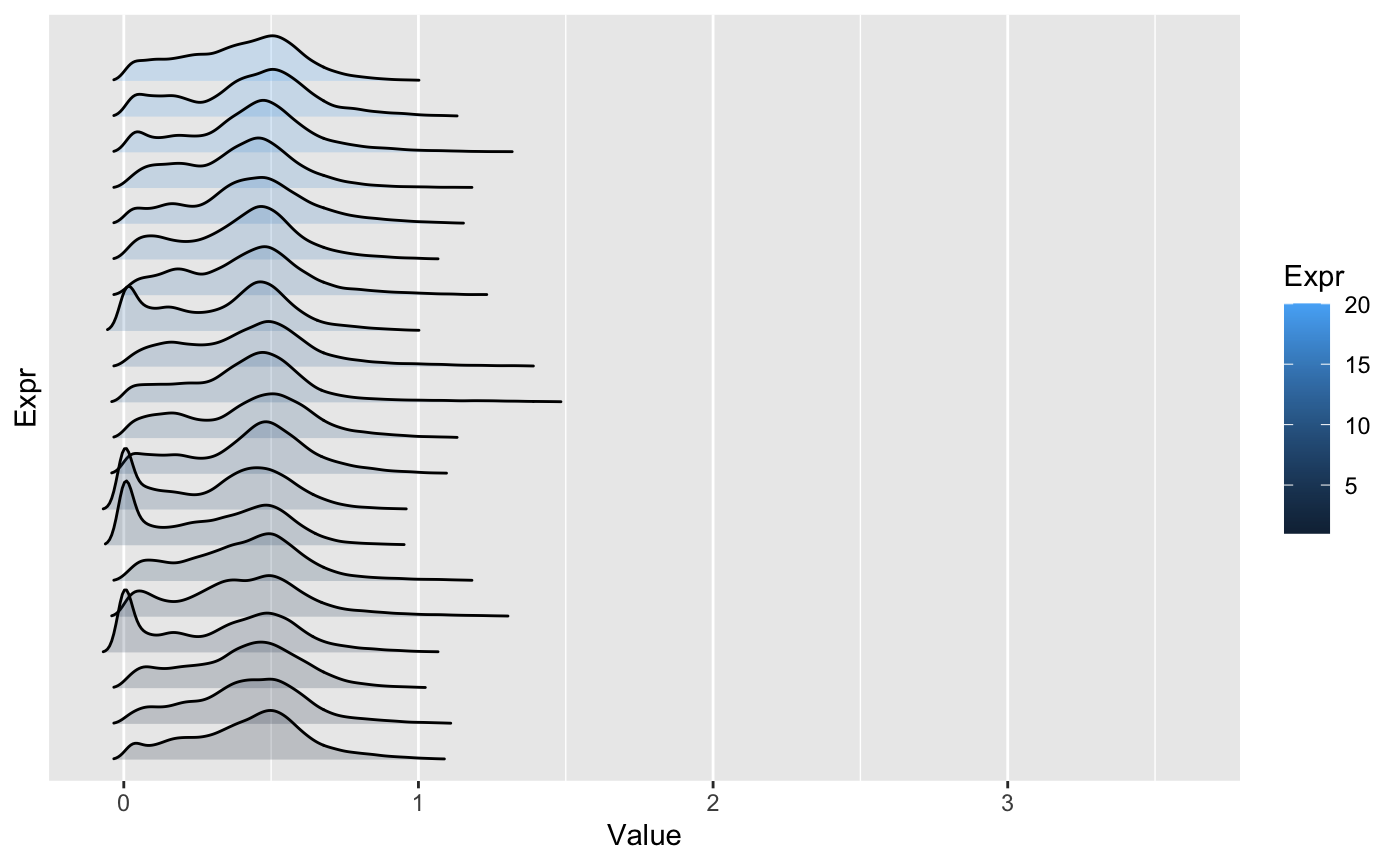}}
\caption{Sensitivity of posterior distributions of \gls{sv} parameters to their corresponding priors. We draw 20000 MCMC samples after discarding the 10000 samples as burn-ins. (a) We fix $b_{\mu}=0$ and test $B_{\mu}\in\{1,10\}$. (b) We fix $b_{\phi}=2.5$ and test $a_{\phi}\in \{20,10,2.5\}$, which corresponds to $\mathbb{E}(\phi_{ks}^g)=a_{\phi}/(a_{\phi}+b_{\phi})\in\{0.89,0.8,0.5\}$. (c) We test $B_{\sigma}\in\{0.5,1,10\}$.}
\label{fig:PriorSensSV}
\end{figure}

For the \gls{sv} process, the mean of logarithmic \gls{icn} amplitude, $\mu_{ks}^g\sim \mathcal{N}(b_{\mu},B_{\mu})$, we fix $b_{\mu}=0$ and test $B_{\mu}\in\{1,10\}$. Note that though $B_{\mu}$ corresponds to very high amplitude at the exponential level and covers a wide range of possible values of $\mu_{ks}^g$. The \gls{ar}(1) coefficient $\phi_{ks}^g$ follows a prior distribution, $\left(\phi_{ks}^g+1\right)/2\sim\text{Beta}\left(a_{\phi},b_{\phi}\right)$. We fix $b_{\phi}=2.5$ and test $a_{\phi}\in \{20,10,2.5\}$, which corresponds to $\mathbb{E}(\phi_{ks}^g)=a_{\phi}/(a_{\phi}+b_{\phi})\in\{0.89,0.8,0.5\}$. For the variance of \gls{sv}, $\delta^2_{ksg}\sim\text{Gamma}(0.5,0.5B_{\sigma})$, we test $B_{\sigma}\in\{0.5,1,10\}$. \Cref{fig:PriorSensSV} shows that the ranges of posterior distributions remains stable when the prior parameter values vary. This indicates the indeterminacy of hyper-parameters has a limited impact of the corresponding posteriors. As indicated by the \gls{hcp} data, we set $B_{\mu}=1$, $a_{\phi}=20$, and $B_{\sigma}=0.5$.

The regional-specific variances follow a weakly informative inverse Gamma distribution. We set the shape parameter $c_{\sigma}=2$ and the rate parameter $d_{\sigma}=1/\widehat{\text{Var}}\left(\text{vec}\left(\bY\right)\right)$ to match the mean of this prior distribution with the empirical variance $\widehat{\text{Var}}\left(\text{vec}\left(\bY\right)\right)$. 

To select a proper number of latent factors for the multi-subject model, we run the \gls{bicnet} model under the single-subject setting where the sampling of group inclusion probability, $\boldsymbol{\Pi}_0=\left(\pi_{nk}\right)$, is not performed. We calculate \gls{aic}, \gls{bic}, and \gls{dic} on $\widehat{K}\in\{2,3,\dots,29,30\}$. The \gls{dic} is a Bayesian generalization of \gls{aic}. Similar to \gls{aic}, \gls{dic} is also an asymptotic estimate of the Kullback-Leibler divergence. We store a chain of 10000 samples after discarding the first 20000 samples as burn-in. For $\widehat{K}\in[14,24]$, we run 4 extra MCMC chains with different initial values, and calculate the mean and standard deviation for each $\widehat{K}$. We use the posterior means as estimators for unknown parameters.

\begin{figure}
\centering
\subfloat[]{\includegraphics[width=.45\textwidth]{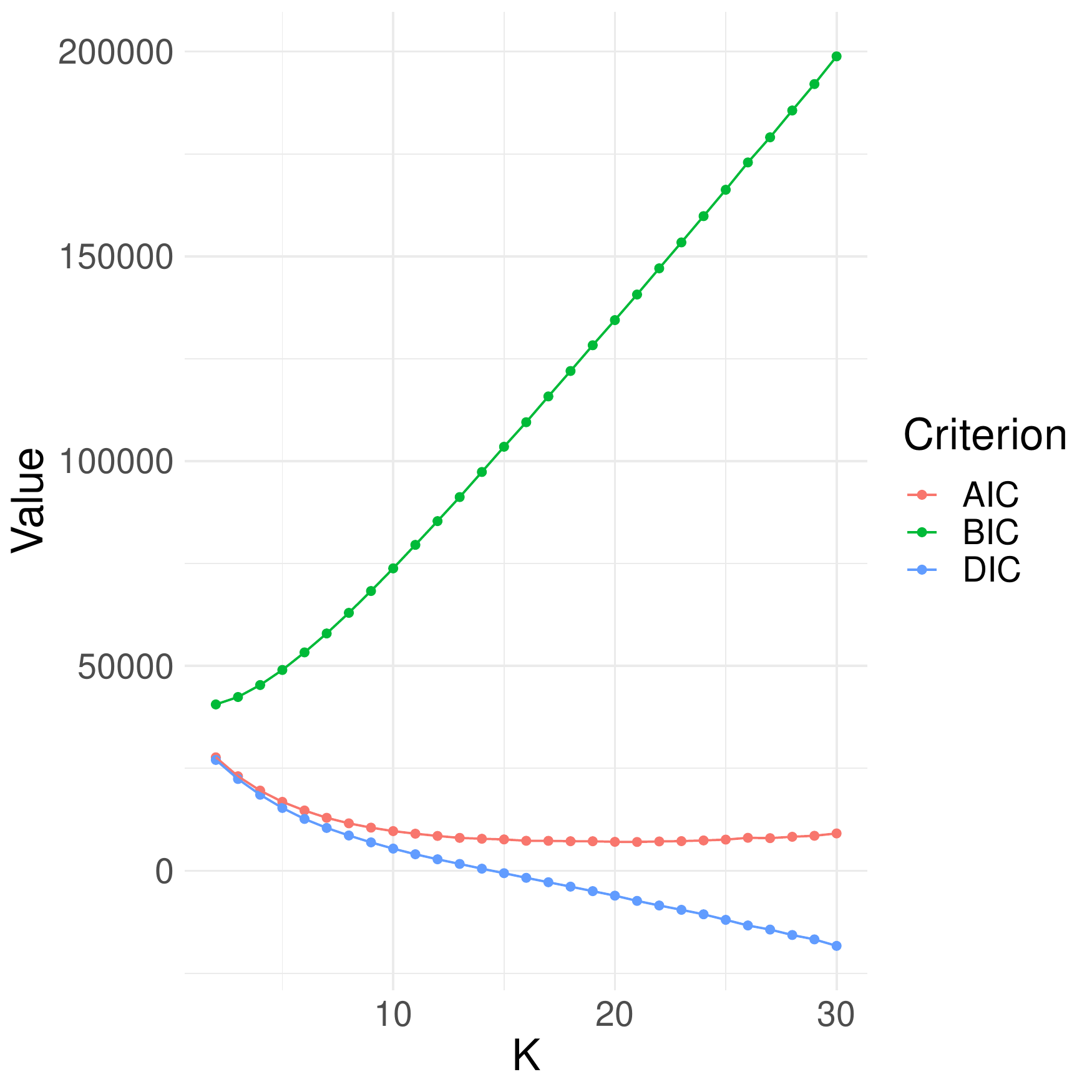}}
\subfloat[]{\includegraphics[width=.45\textwidth]{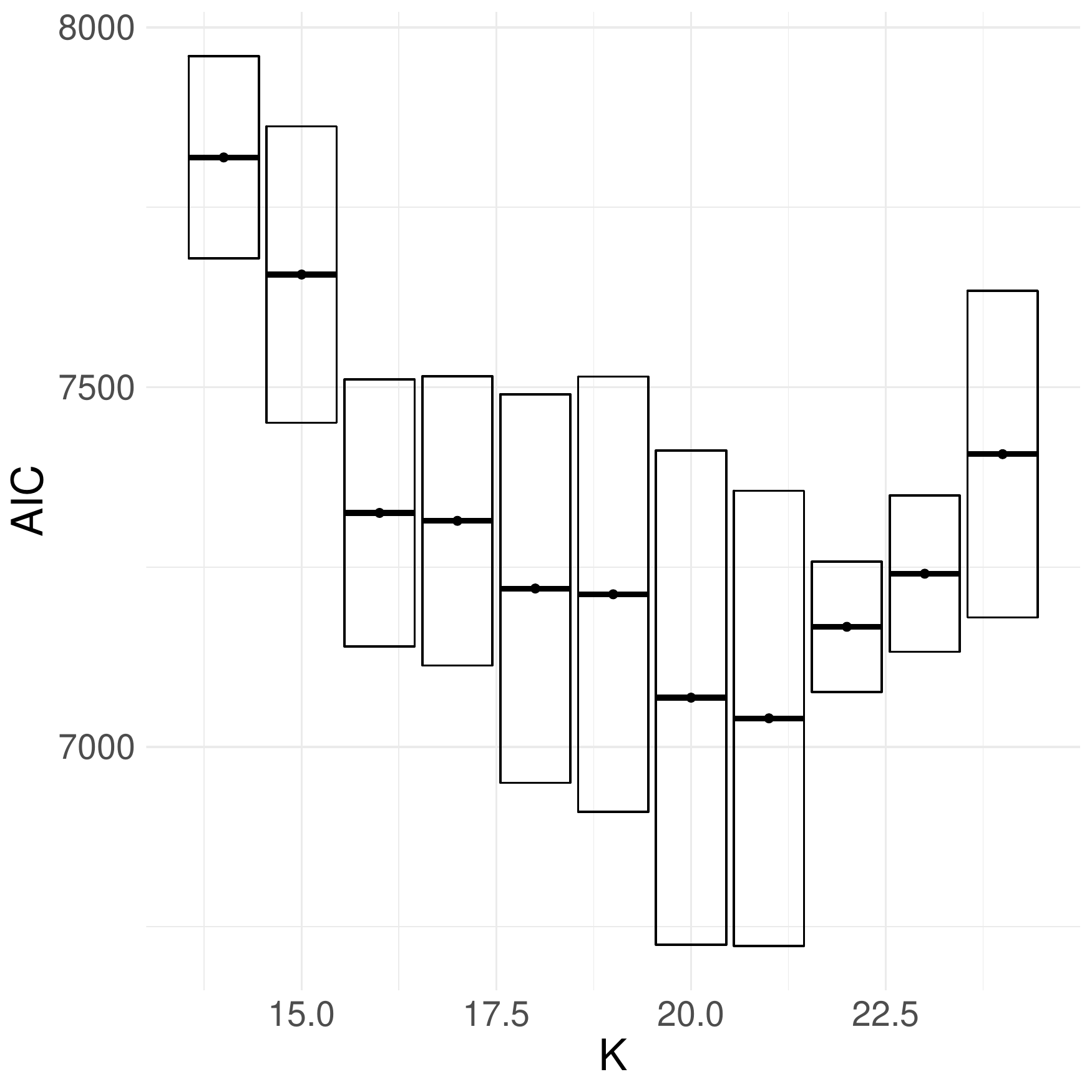}}
\caption{Selecting number of \glspl{icn} with \gls{aic}, \gls{bic}, and \gls{dic} on $\widehat{K}\in\{2,3,\dots,29,30\}$. Note that \gls{aic} and \gls{bic} are calculated with the posterior mean estimator. (a) \gls{aic} decreases first and then remains relatively stable since $\widehat{K}=14$. \gls{bic} keeps increasing and thus favors the simplest model with $\widehat{K}=2$. \gls{dic} continues to decrease and thus prefers the most complex model with $\widehat{K}=30$. However, large negative \gls{dic} might imply the MCMC chains do not converge well, or the distribution is over-dispersed. (b) We zoom in the \gls{aic} for $\widehat{K}$ from $14$ to $24$. For each $\widehat{K}$, 5 MCMC chains with different initial values were ran and then the means and standard deviations were calculated. It is reasonable to choose $\widehat{K}\in[16,24]$ since the \gls{aic} values do not have significant differences. We choose $\widehat{K}=20$ because its \gls{aic} value is relatively smaller than those of $\widehat{K}<20$ and it has lower model complexity than those of $\widehat{K}>20$.}
\label{fig:RankSelection}
\end{figure}

From \cref{fig:RankSelection}(a), it is apparent that \gls{bic} continues increasing and thus favors the simplest model with $\widehat{K}=2$. On the other hand, \gls{dic} keeps decreasing and thus prefers the most complex model with $\widehat{K}=30$. In contrast, \gls{aic} first decreases and then remains relatively stable for $\widehat{K}\geq 14$. Based on \cref{fig:RankSelection}(b), we chose $\widehat{K}=20$ because it has a relatively lower \gls{aic} value than those of $\widehat{K}<20$ and a lower model complexity than those of $\widehat{K}>20$. For the MCMC sampling, a chain of 10000 samplers where stored after discarding the first 40000 samples as burn-in without thinning.
To obtain interpretable group-level spatial maps, we threshold $\boldsymbol{\Pi}_0$ at $0.999$ in this study. For group \gls{ica} results, the spatial maps are transformed into z-scores, and the z-scores were thresholded.

\subsection{Results}

\subsubsection{Findings on both prevalent and highly exclusive regions across group-level spatial maps}

\begin{figure}[htbp]
\centering
\includegraphics[width=.9\textwidth]{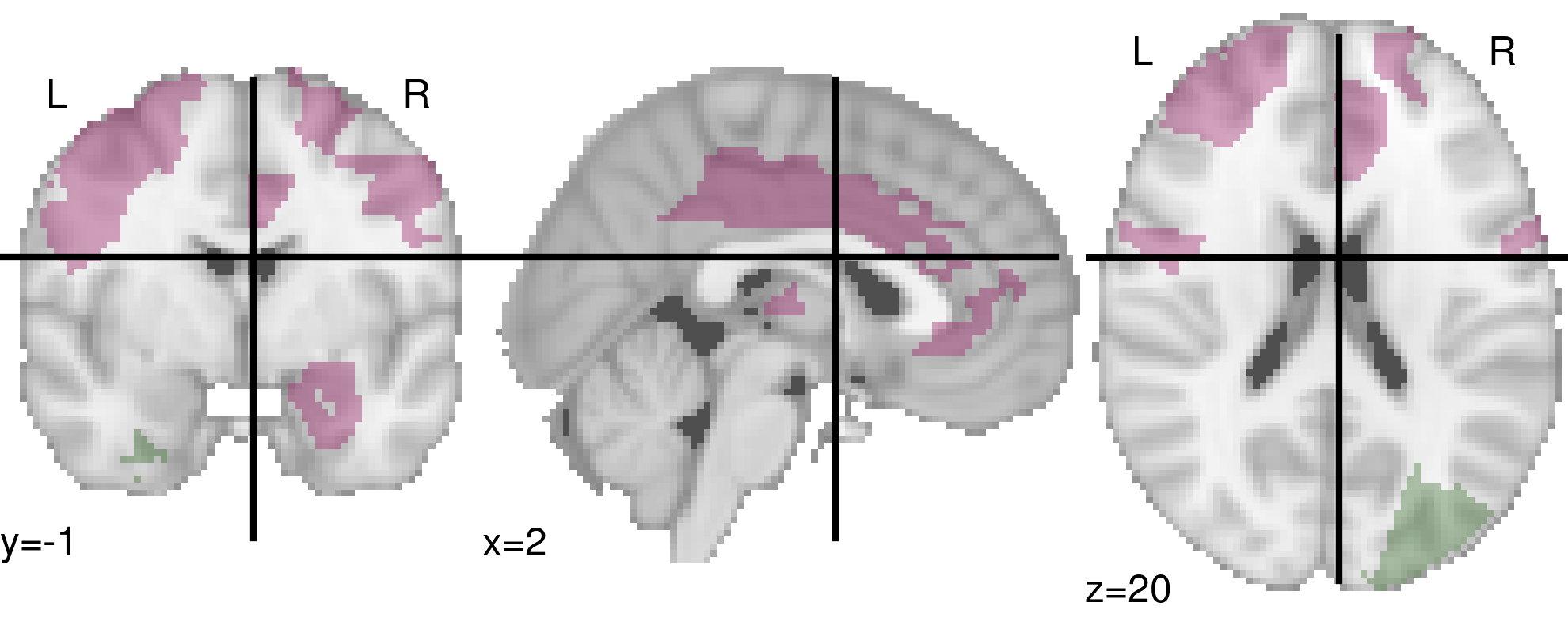}
\caption{Exclusive regions in pink consist of mainly the \gls{sfg}, which is related to many high-order cognitive function. Prevalent regions locate in the occipital lobe.}
\label{fig:FacPerNode}
\end{figure}

One interesting observation is that while there are regions that appear as nodes in many \glspl{icn}, some regions are more exclusive to specific \glspl{icn}. Both prevalent and exclusive regions are presented in \cref{fig:FacPerNode}. For the remainder of this paper, an \gls{roi} is said to belong to a particular \gls{icn} at the group level if the inclusion probability is greater than 0.999.

The exclusive regions in pink are defined to be those \glspl{roi} that only appear in one group spatial map. It is noteworthy that these exclusive regions are mainly located in the \gls{sfg} symmetrically. The low inclusion probability suggests that the superior frontal region is not as active as other regions, which might arise due to the simplicity of the language tasks. It also suggests that the high-order cognition processes related to \gls{sfg} have high between-subjects variability.

The prevalent regions in green consist of \glspl{roi} that appear in more than seven group spatial maps. It is mainly located in the occipital lobe, suggesting the occipital lobe is constantly active under both resting state and language tasks. Under resting state, participants were asked to fix their eyes on a bright cross, contributing to this activation. Though language tasks did not explicitly involve visual information, the occipital lobes still played an important role.

\subsubsection{Distinct language spatial maps at the group level}

\begin{figure}[htbp]
\centering
\subfloat[Group spatial map of \gls{icn} 3.]{\includegraphics[width=.9\textwidth]{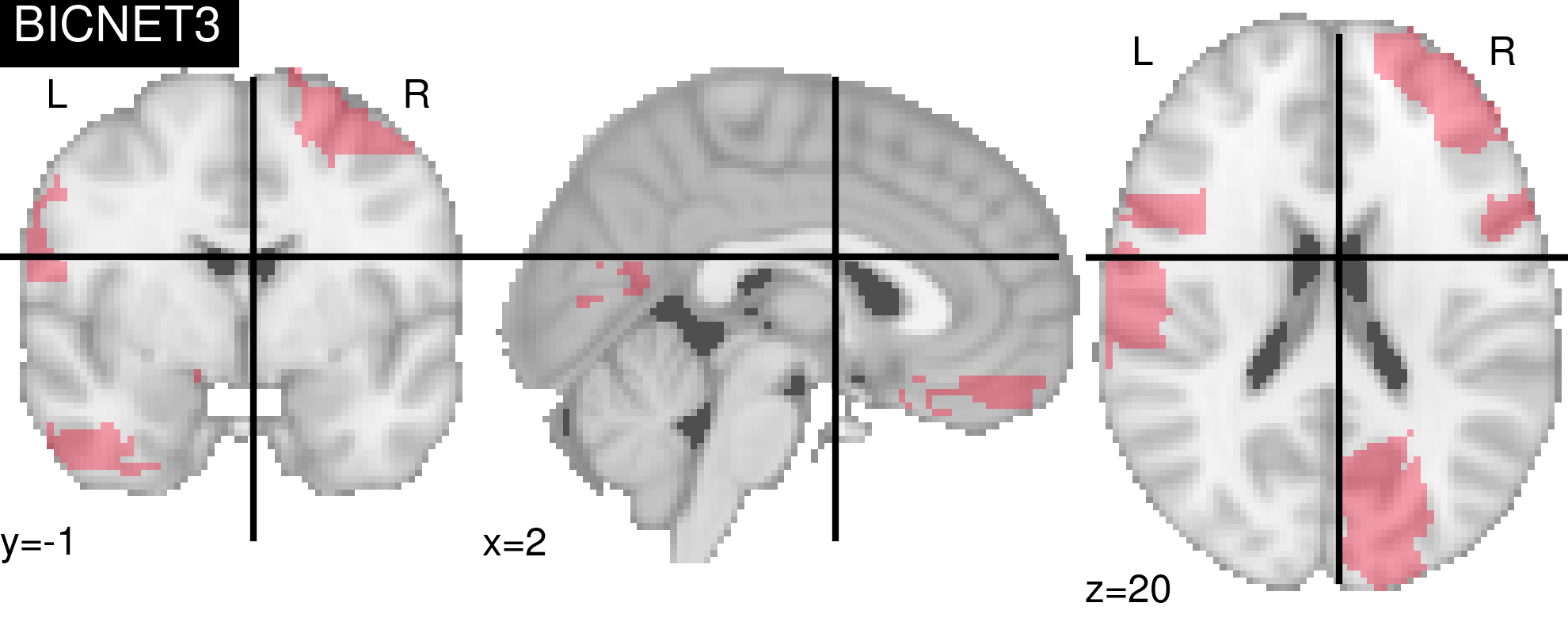}}\\
\subfloat[Group spatial map of \gls{icn} 6.]{\includegraphics[width=.9\textwidth]{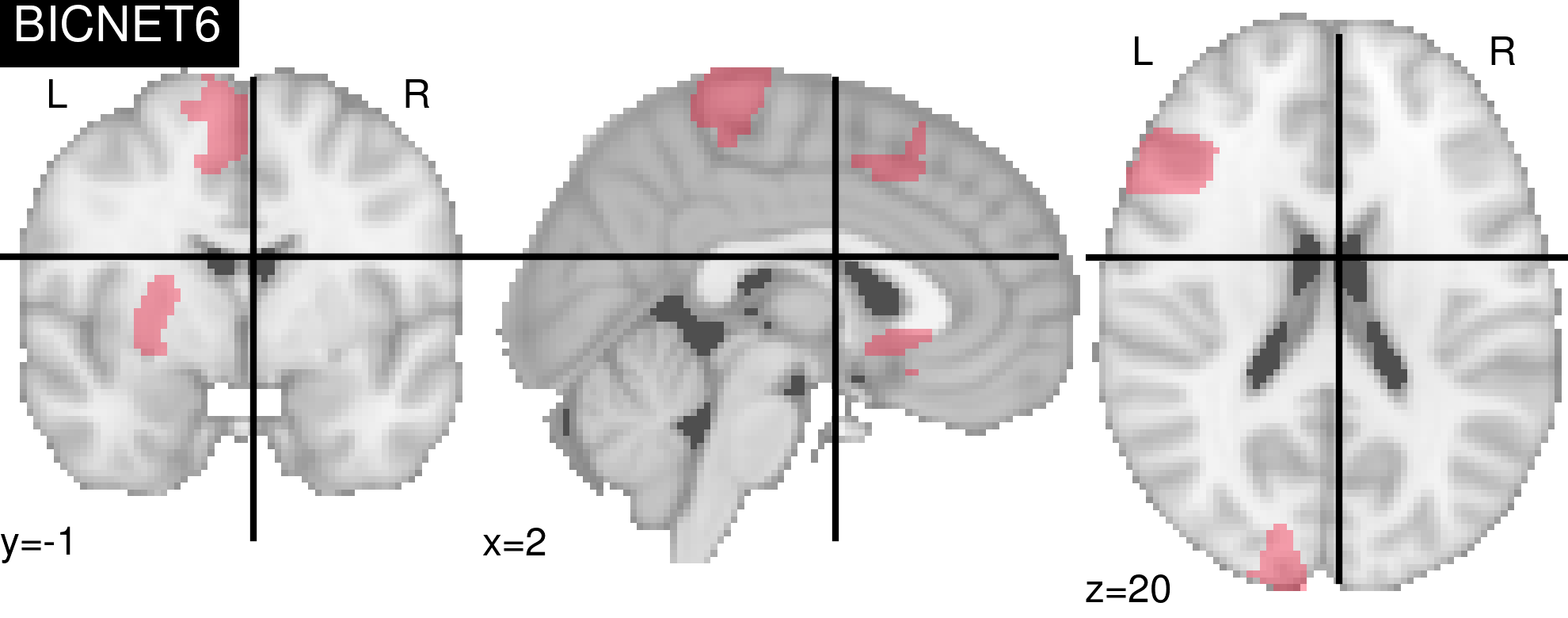}}
\caption{Group-level spatial maps, measured by inclusion probability, that have significant relationships with language tasks' accuracy. (a) Group spatial map of \gls{icn} 3 mainly consists of Broca's area, right frontal lobe, basal ganglia, and the occipital lobe. (b) Group spatial map of \gls{icn} 6 mainly involves the left frontal lobe and left occipital lobe.}
\label{fig:GrpIncldLangAcc}
\end{figure}

\begin{figure}[htbp]
\centering
\subfloat[Group spatial map of \gls{icn} 2, related to story difficulty level with $p=0.040$.]{\includegraphics[width=.9\textwidth]{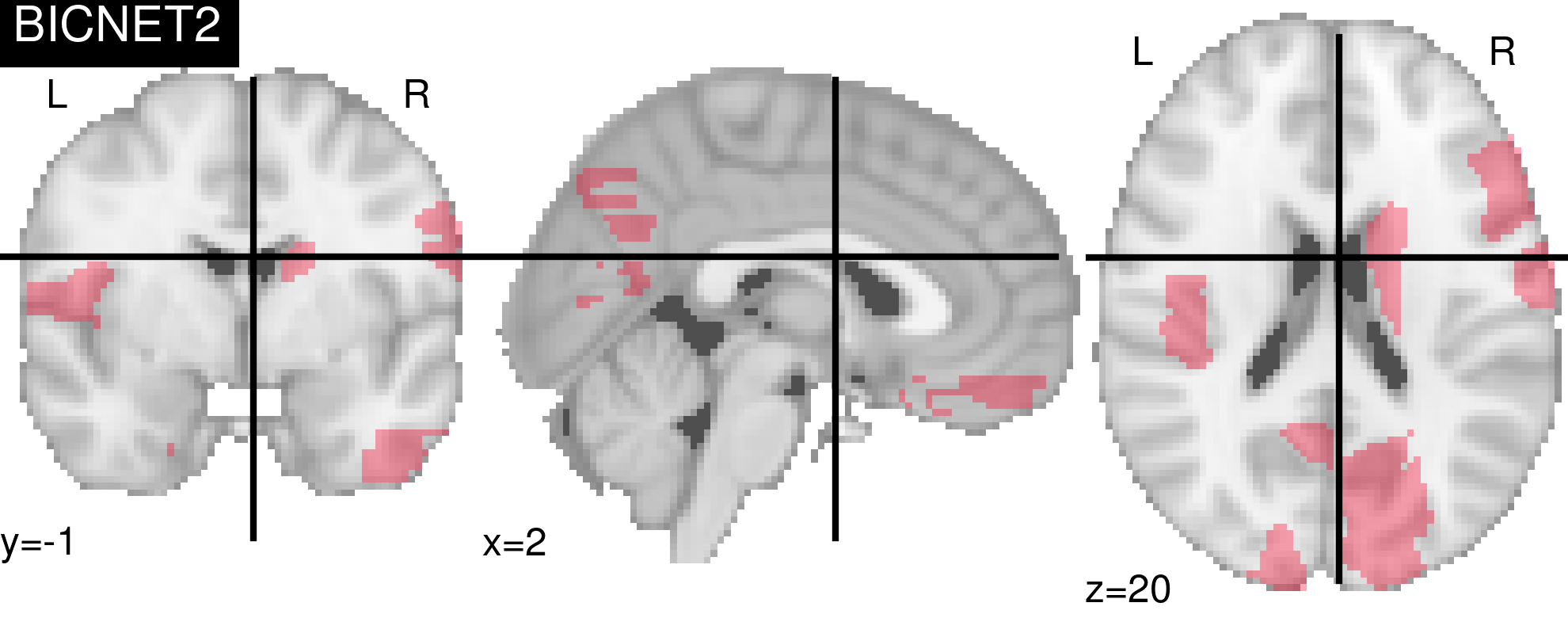}}\\
\subfloat[Group spatial map of \gls{icn} 19, related to story difficulty level with $p=0.011$.]{\includegraphics[width=.9\textwidth]{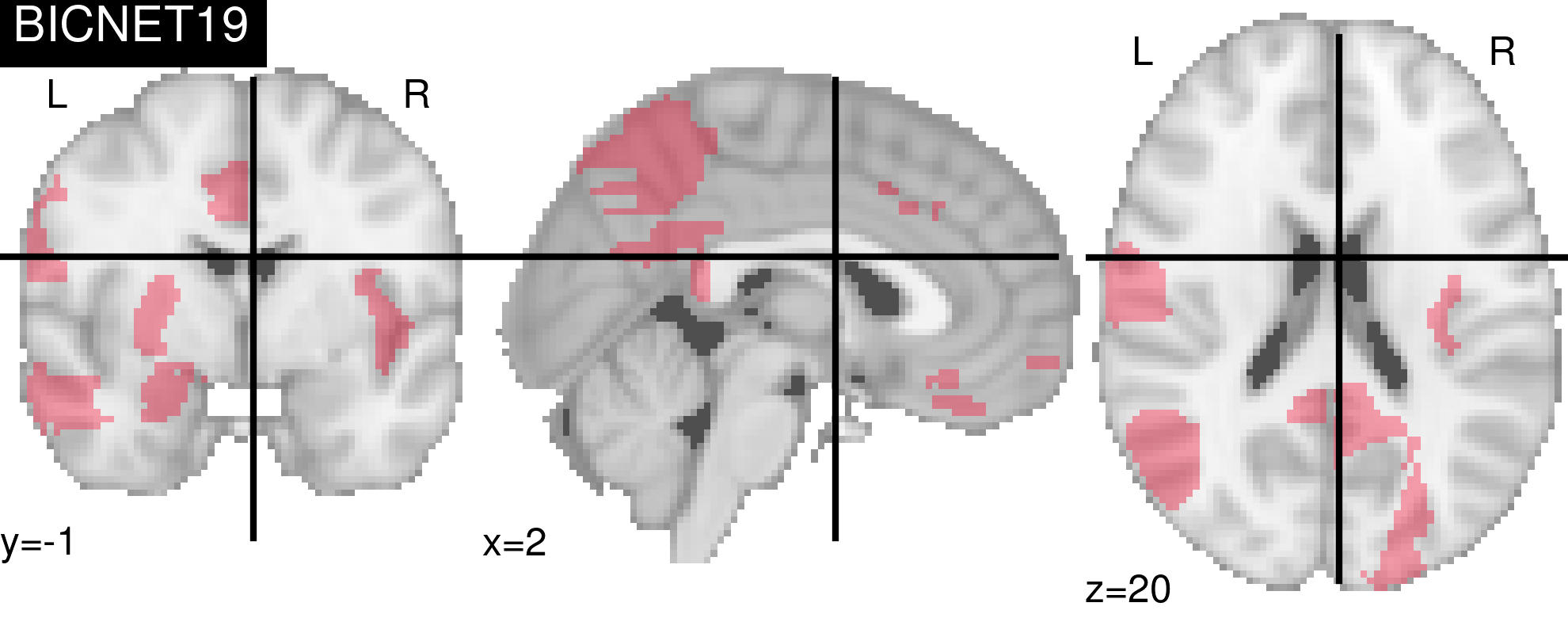}}
\caption{Group-level spatial maps, measured by inclusion probability, that have significant relationships with story tasks' difficulty levels. (a) Group spatial map of \gls{icn} 2 has main contributions of the right hemisphere and left frontal lobe. (b) Group spatial map of \gls{icn} 19 receives contributions from mainly the left hemisphere.}
\label{fig:GrpIncldStoryDiff}
\end{figure}

\begin{figure}[htbp]
\centering
\subfloat[Group spatial map of \gls{icn} 14, related to math difficulty level with $p=0.040$.]{\includegraphics[width=.9\textwidth]{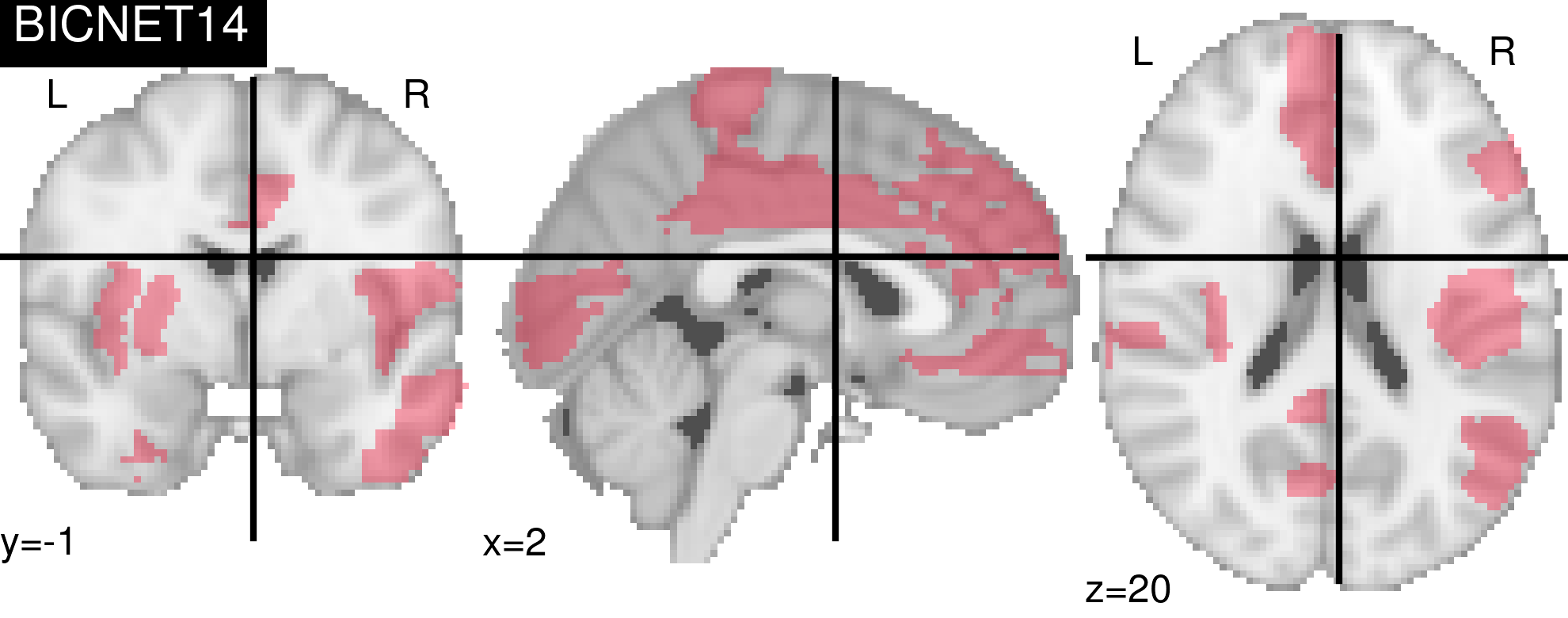}}\\
\subfloat[Group spatial map of \gls{icn} 17, related to math difficulty level with $p=0.011$.]{\includegraphics[width=.9\textwidth]{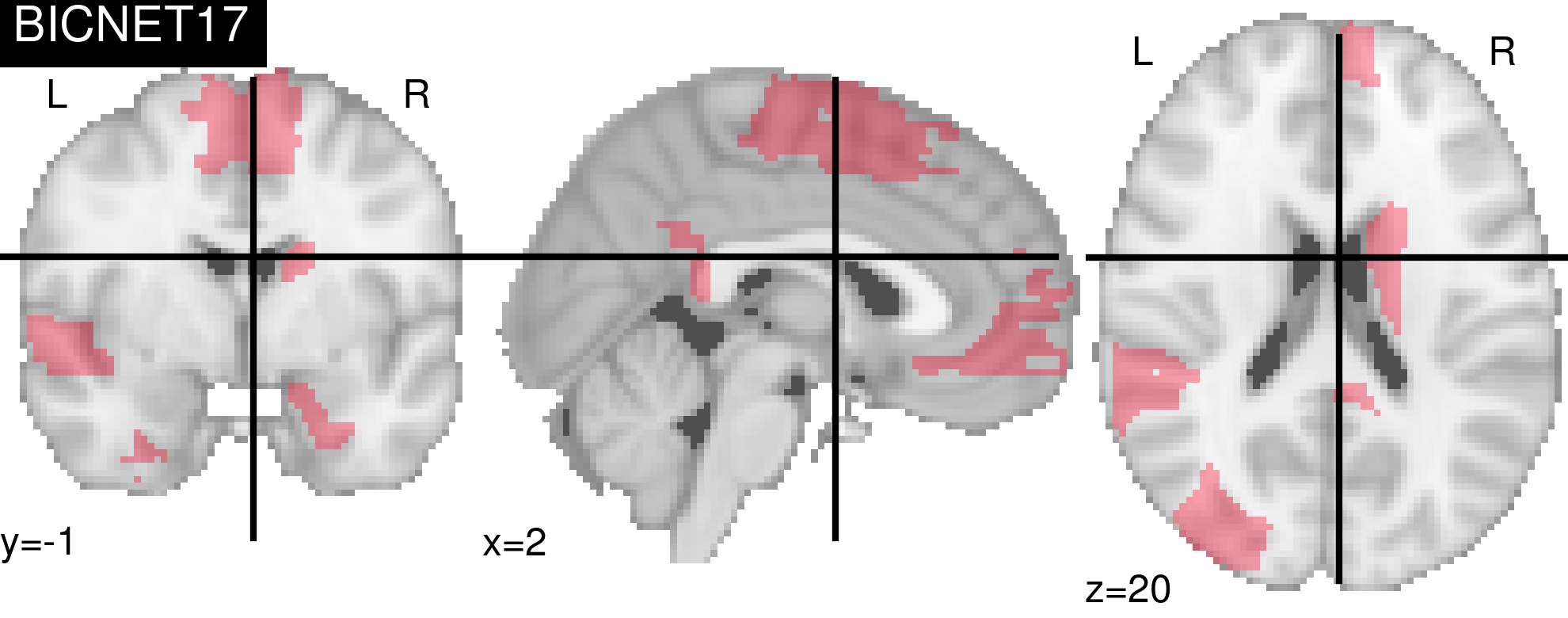}}
\caption{Group-level spatial maps, measured by inclusion probability, that have significant relationships with math tasks' difficulty levels.}
\label{fig:GrpIncldMathDiff}
\end{figure}

As shown in \cref{fig:GrpIncldLangAcc,fig:GrpIncldStoryDiff,fig:GrpIncldMathDiff}, the proposed \gls{bicnet} yielded very interesting results that show the association between behavioral measures and changes in the amplitudes of \glspl{icn} between language task and resting state. 


\Cref{fig:GrpIncldLangAcc} shows group spatial maps of \glspl{icn} 3 and 6 that are correlated to language task accuracy. The group spatial map of \gls{icn} 3 mainly consists of the language network, right frontal lobe, and occipital lobe. It indicates that though the language function is closely related to the left hemisphere, the right frontal lobe is also involved in this process and probably helps with understanding and interpretation. \cite{Voets2006} also reports right frontal lobe activation when patients' left hemispheres are injured. \gls{icn} 6 is mainly the language network on the left hemisphere. 

These observations emphasize the fundamental role of language networks in language processing. However, language processing involves sensory input, motor output, attention, and a short-term memory system. To further understand language processing, it is necessary to analyze them in a finer functional resolution.

Group spatial maps of \gls{icn} 2 and 19 show interesting asymmetric patterns related to the story difficulty level. \Gls{icn} 2 has main contributions from the right hemisphere. On the contrary, the group spatial map of \gls{icn} 19 focuses more on the left hemisphere, which involves the language network. It indicates that these two \glspl{icn} are related to language processing and high-level cognitive process. \cite{Corballis2014} demonstrates that vision and attention are biased toward the right hemisphere while language and internal thought are biased towards the left hemisphere. While the left brain is credited with language, the right brain helps us understand the context and tone. Also, both spatial maps involve the parietal lobe, which suggests that the parietal lobe also plays an essential role in language processing.

Similarly, \glspl{icn} related to math tasks' difficulty levels are also asymmetric with group spatial map of \gls{icn} 14 focus in the right hemisphere and group spatial map of \gls{icn} 17 focus in the left hemisphere. Compared to the \glspl{icn} related to story tasks' difficulty levels, these \glspl{icn} are mainly located in the frontal and parietal lobe.

Overall, group spatial maps related to behavioral measures tend to be asymmetric while focus on some specific regions. This phenomenon indicates the task execution requires functional integration of different regions while some specific regions still play a dominant role.

\subsubsection{Comparison to group ICA spatial maps estimated by GIFT}

We also estimated 20 group \gls{ica} spatial maps using the \gls{gift} algorithm. The spatial maps estimated by \gls{gift} are converted to the corresponding one-sample $t$-statistics and thresholded at $t=3.34$ ($\text{df}=199$, $p<0.001$). As shown in \cref{eq:jaccard}, we use Jaccard similarity to measure the overlap between the spatial maps estimated by group \gls{ica} and \gls{bicnet}, and thus obtain a one-to-one mapping between them. 
\begin{equation}
J(S_i^B,S_j^G) = \frac{\left|S_i^B\cap S_j^G\right|}{\left|S_i^B\cup S_j^G\right|},
\label{eq:jaccard}
\end{equation}
where $S_i^B$ is the set of \glspl{roi} in \gls{bicnet} group spatial map $i$ and $S_j^G$ is the set of \glspl{roi} in \gls{gift} group spatial map $j$. The average Jaccard similarity is $0.0672$ with a standard deviation of $0.0408$. 

\begin{figure}[htbp]
\centering
\subfloat{\includegraphics[width=0.45\textwidth]{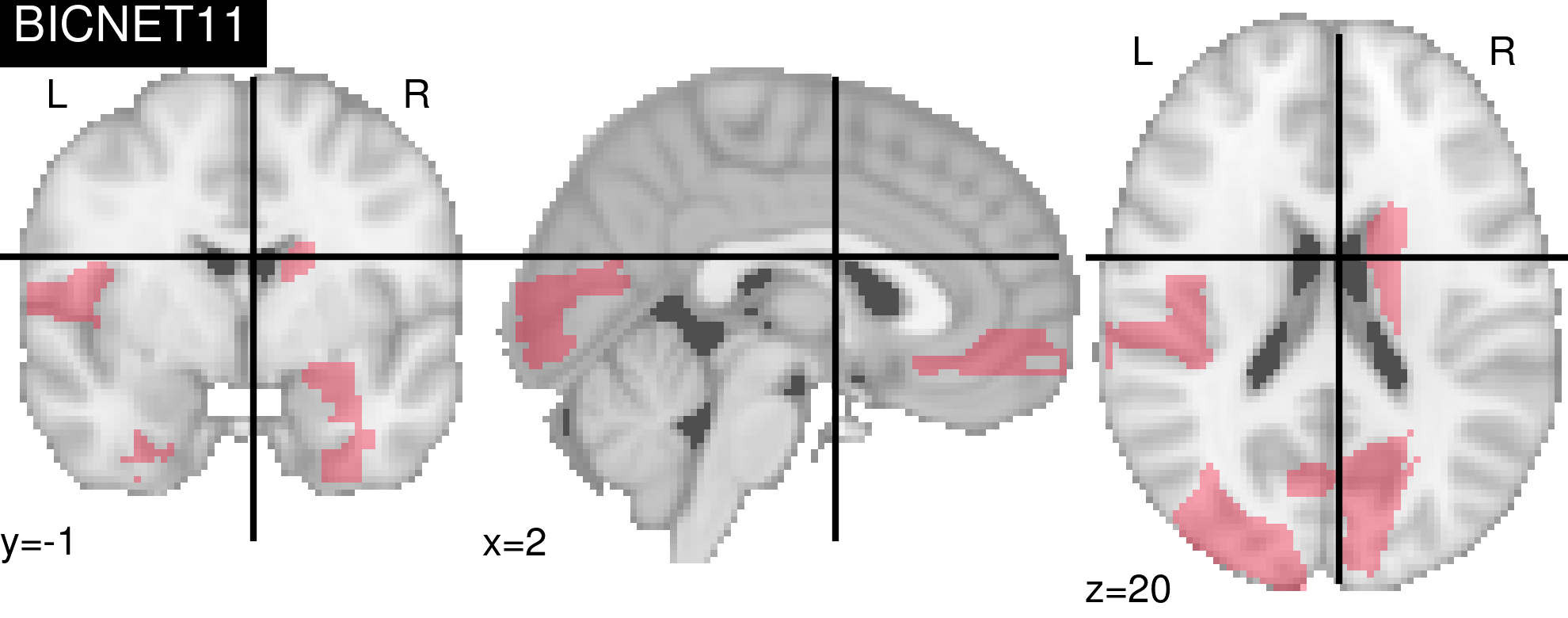}}\hspace{+0.1in}
\subfloat{\includegraphics[width=0.45\textwidth]{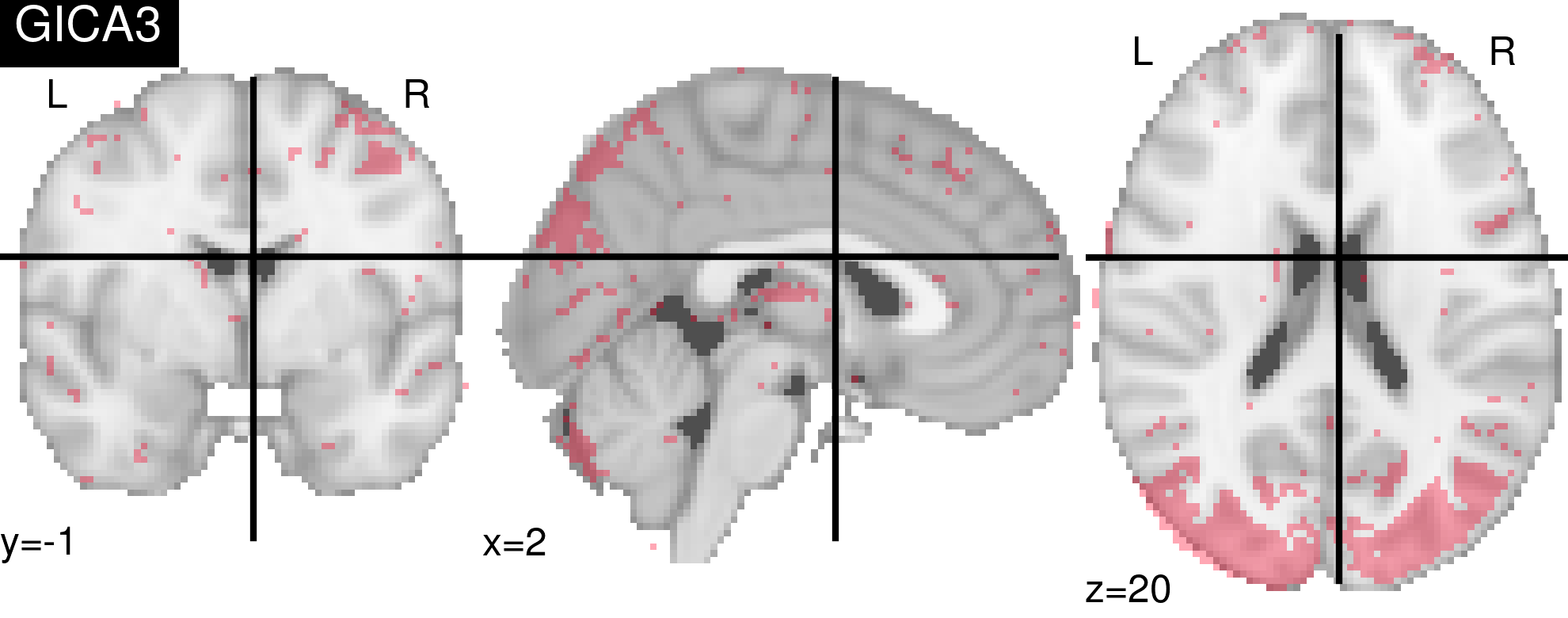}}\\
\subfloat{\includegraphics[width=0.45\textwidth]{BICNETSpatialMap3.png}}\hspace{+0.1in}
\subfloat{\includegraphics[width=0.45\textwidth]{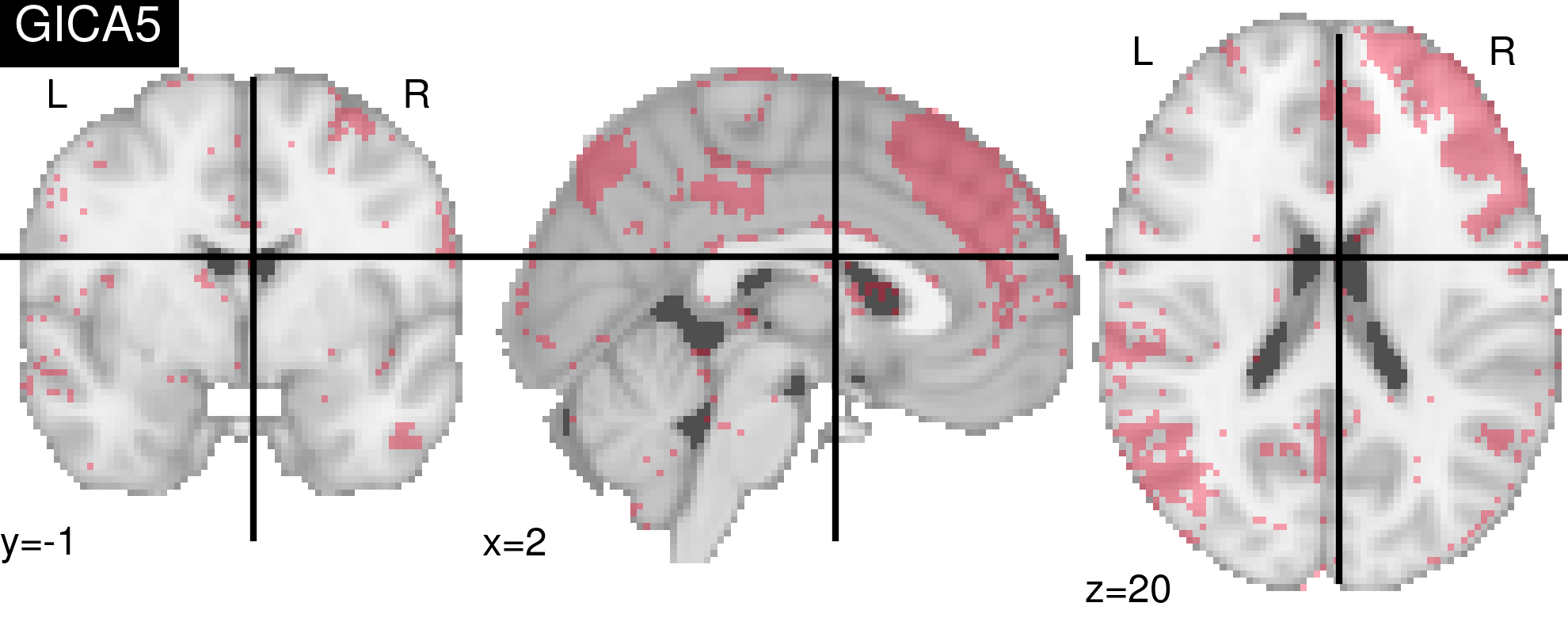}}\\
\subfloat{\includegraphics[width=0.45\textwidth]{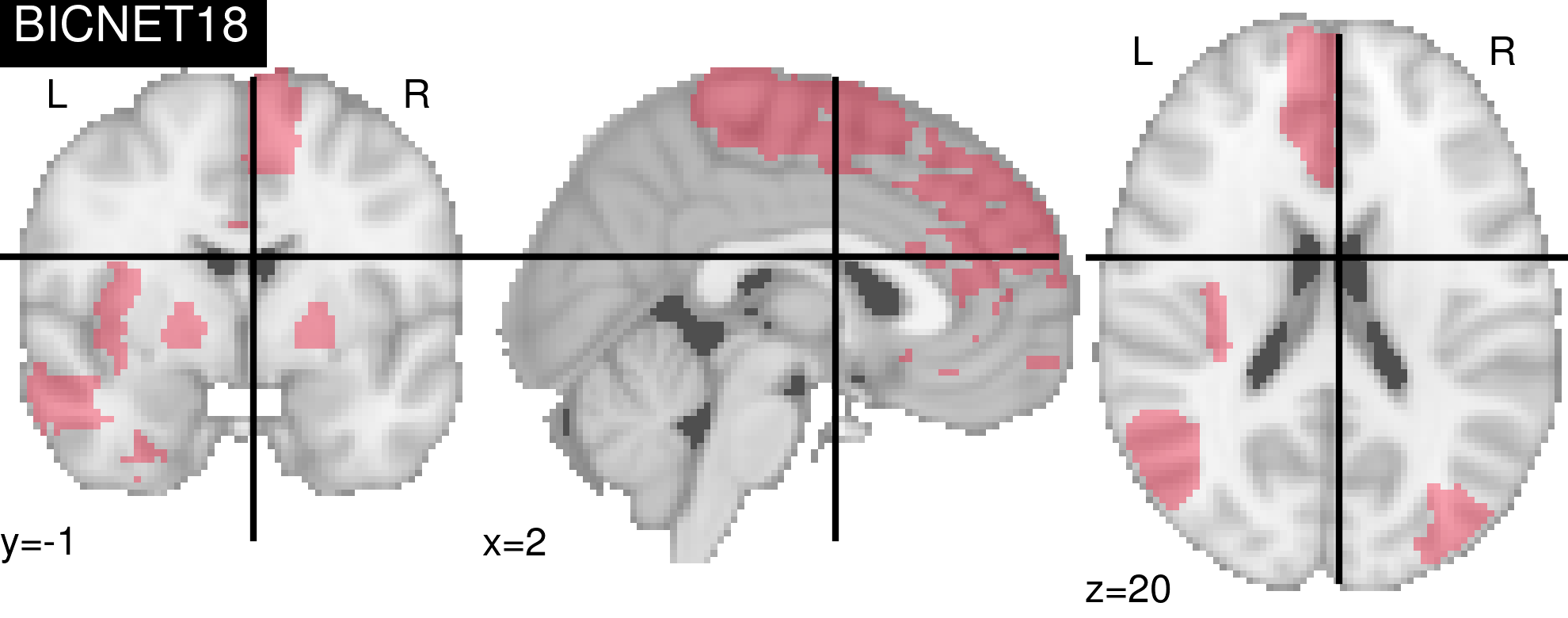}}\hspace{+0.1in}
\subfloat{\includegraphics[width=0.45\textwidth]{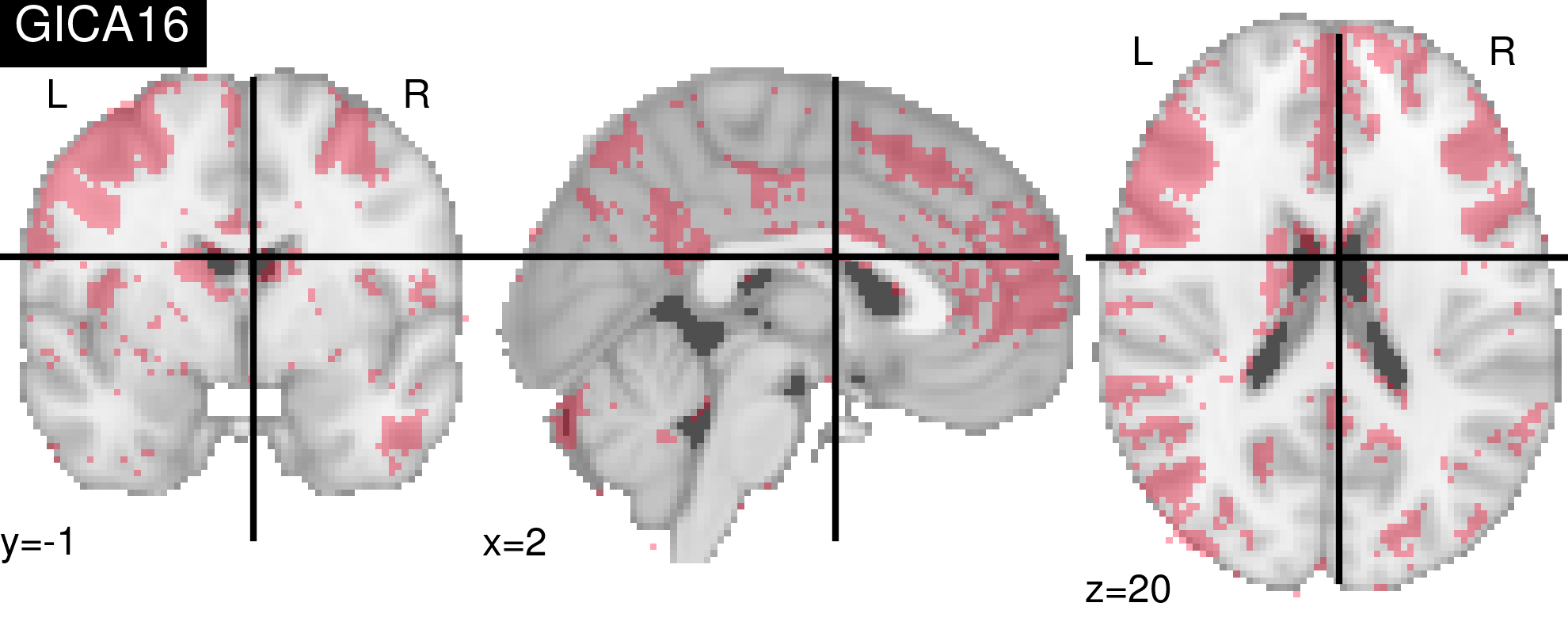}}\\
\subfloat{\includegraphics[width=0.45\textwidth]{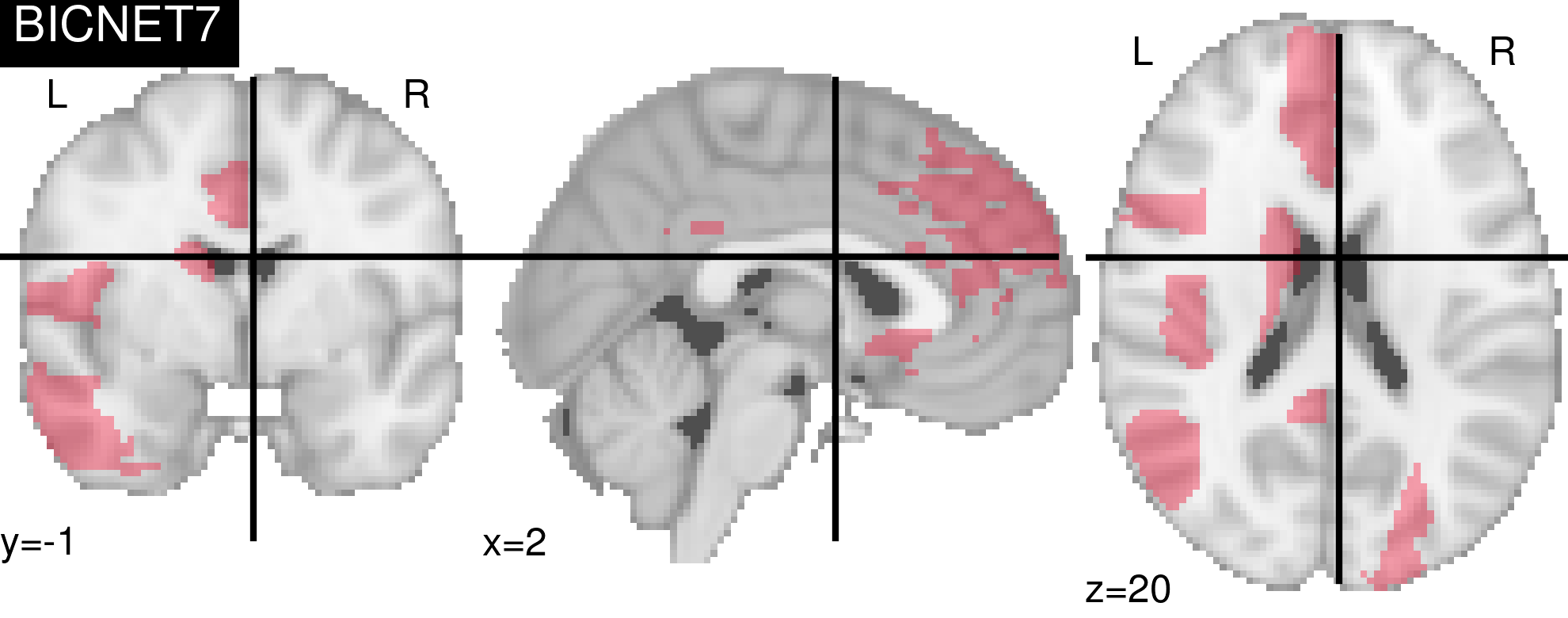}}\hspace{+0.1in}
\subfloat{\includegraphics[width=0.45\textwidth]{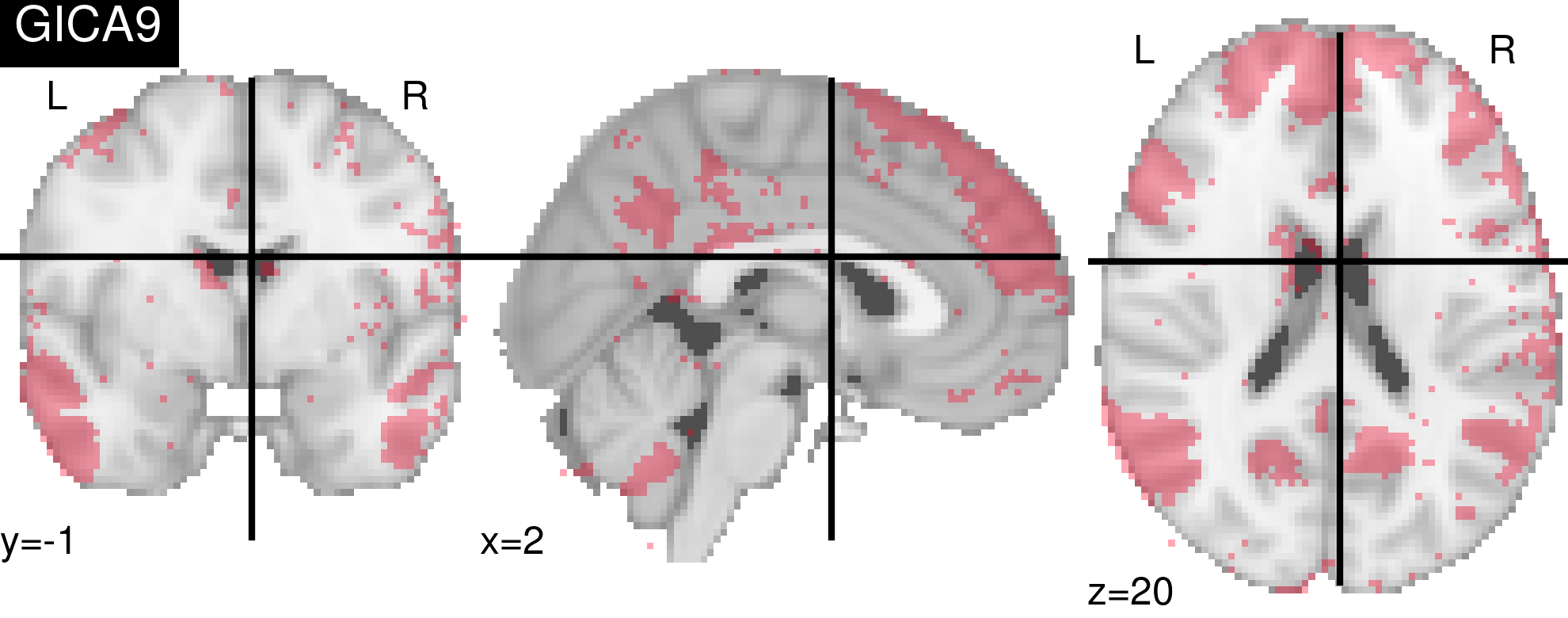}}
\caption{Top four pairs of \gls{bicnet}-\gls{gift} spatial maps with Jaccard similarity higher than $0.0672+2\cdot0.0408$, sorted by Jaccard similarity in a descending order. The left column is the \gls{bicnet} spatial maps and the right column is the \gls{gift} spatial maps.}
\label{fig:bicnetgiftmap}
\end{figure}

\Cref{fig:bicnetgiftmap} presents the four most similar \gls{bicnet}-\gls{gift} spatial map pairs. We notice that \gls{gift} tends to capture symmetric parts in both hemispheres while \gls{bicnet} tends to capture asymmetric integration of \glspl{roi} in two hemispheres. The asymmetry shown in this application and the statistical properties make \gls{bicnet} more suitable to identify functionally meaningful \glspl{icn}.

\subsubsection{Static and Dynamic Amplitudes of individual ICNs}
This analysis clearly demonstrates that the \gls{bicnet} model is able to detect activation states' changes, including no activation, excitation, and inhibition, using either the static or dynamic \gls{icn} amplitudes from resting state to language task performance at the individual level.

\begin{figure}[htbp]
\centering
\subfloat[Proportion of (No) Activation Per ICN]{\includegraphics[width=.45\textwidth]{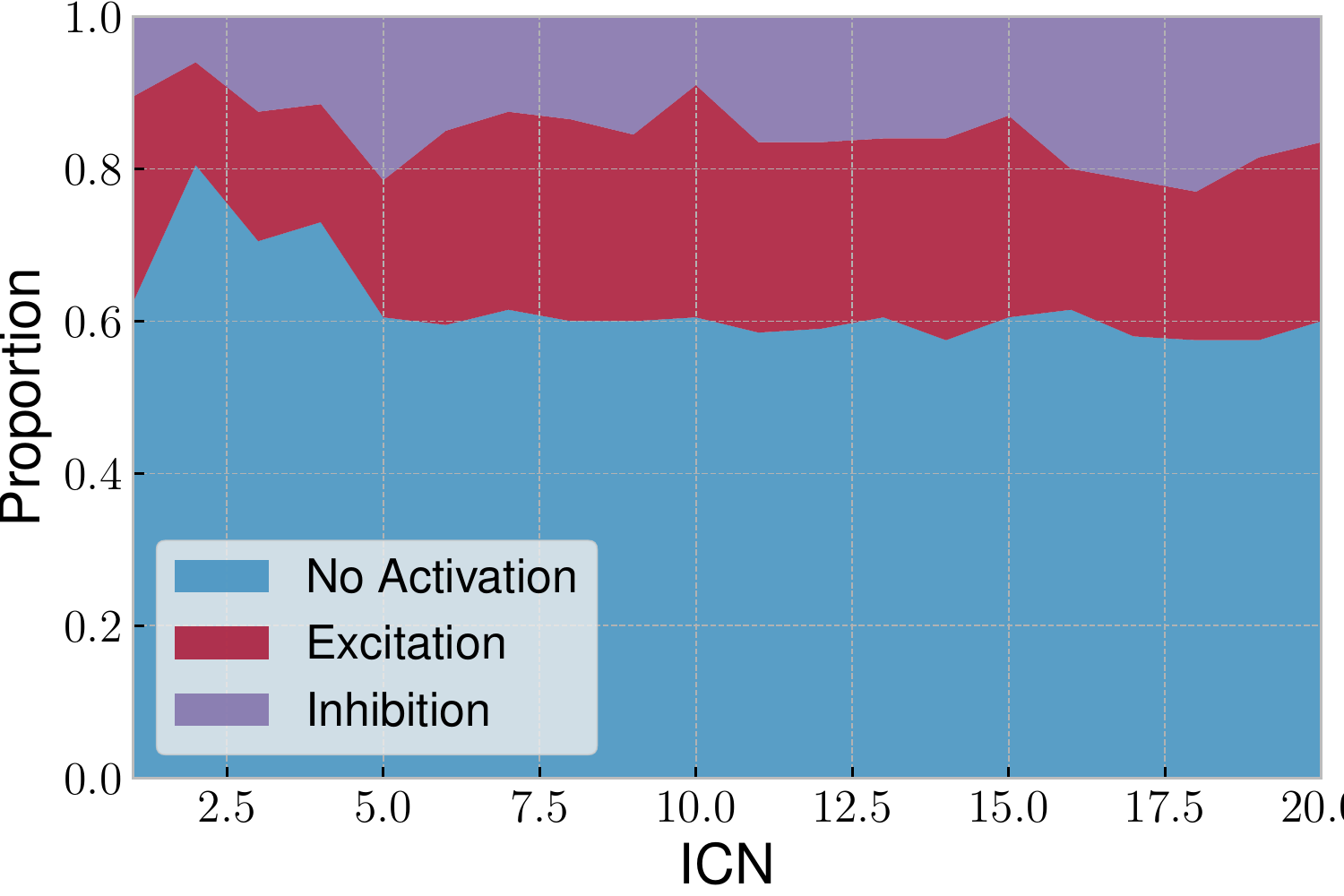}}
\subfloat[Number of Activated ICNs Per Subject]{\includegraphics[width=.45\textwidth]{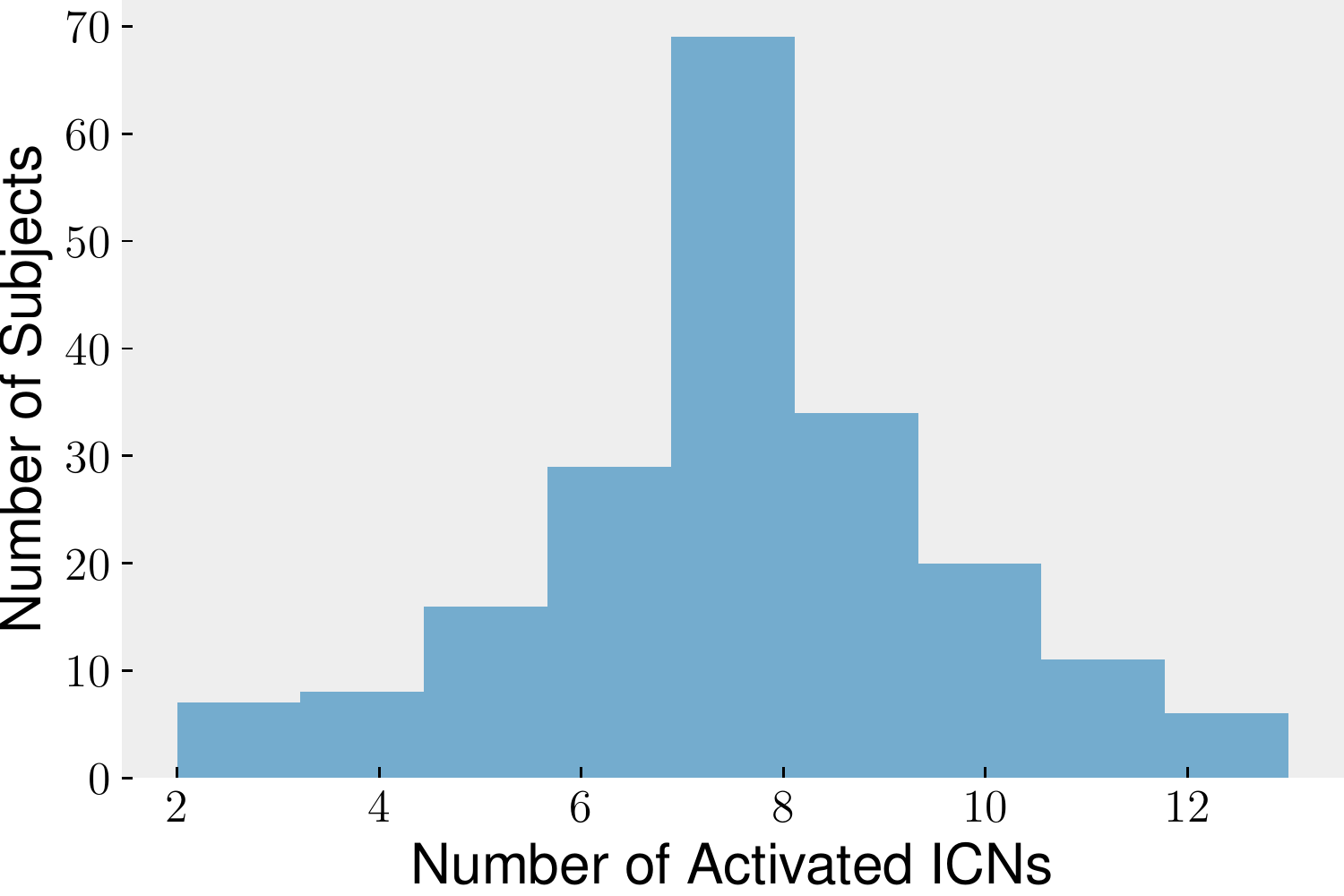}}\\
\subfloat[Activation States of \gls{icn} 4 for the First Four Subjects]{\includegraphics[width=.9\textwidth]{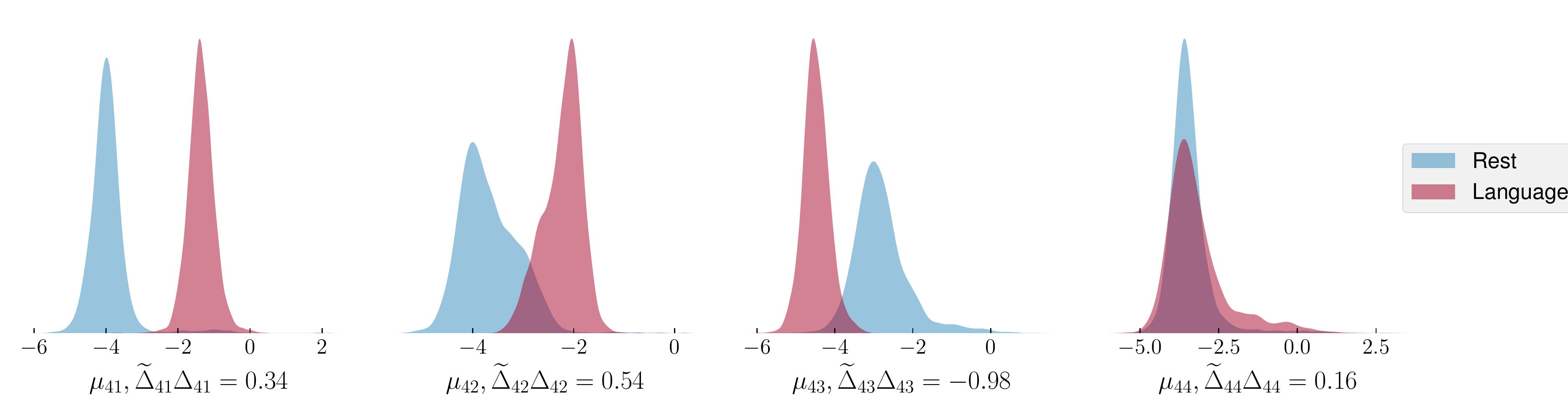}}
\caption{(a) The proportion of each state is the number of subjects with the \gls{icn} in the corresponding state divided by the total number of subjects. The percentage of an \gls{icn} being activated across subjects is relatively low, which indicates high individual variability on \gls{icn} variation. (b) shows that number of activated, including excited and inhibited, \glspl{icn} varies across subjects. (c) As an example, \gls{icn} 4 was activated on subject 1 and 2, inhibited on subject 3 inhibited, and had no impact on subject 4.}
\label{fig:PctAct}
\end{figure}

As shown in \cref{fig:PctAct}(a), an \gls{icn} shows $61.95\%\pm 5.75\%$ of no activation, $22.80\%\pm 4.36\%$ of excitation, and $15.25\%\pm4.25\%$ of inhibition across 200 subjects. The percentage of an \gls{icn} being activated across subjects is relatively low, which indicates high individual variability on \gls{icn} variation. \Cref{fig:PctAct}(b) shows that each subject has at least $40\%$ of \glspl{icn} remain non-activated switching from the resting state to language task execution. \Cref{fig:PctAct}(c) illustrates the inter-subject variability on \gls{icn} activation. For example, \gls{icn} 4 was activated on subject 1 and 2, inhibited on subject 3 inhibited, and had no impact on subject 4.

\begin{figure}[htbp]
\centering
\subfloat[Resting State of Subject 1]{\includegraphics[width=0.3\textwidth]{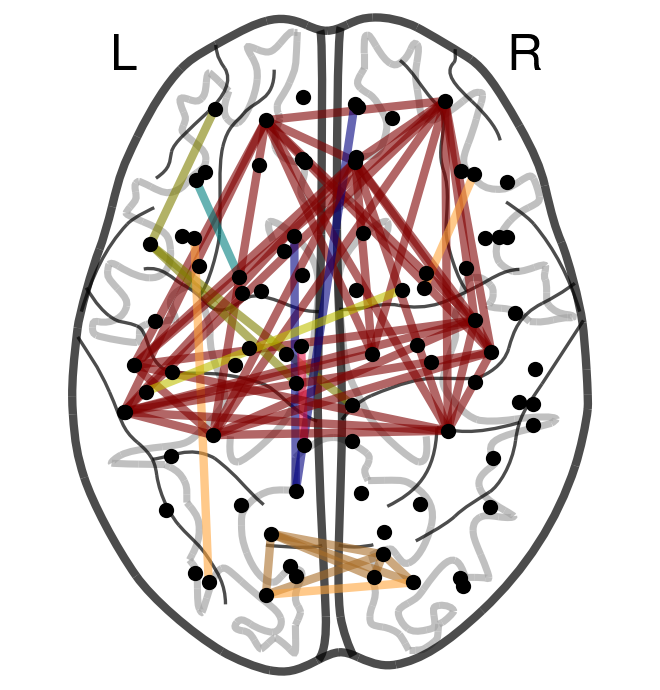}}
\subfloat[Resting State of Subject 2]{\includegraphics[width=0.3\textwidth]{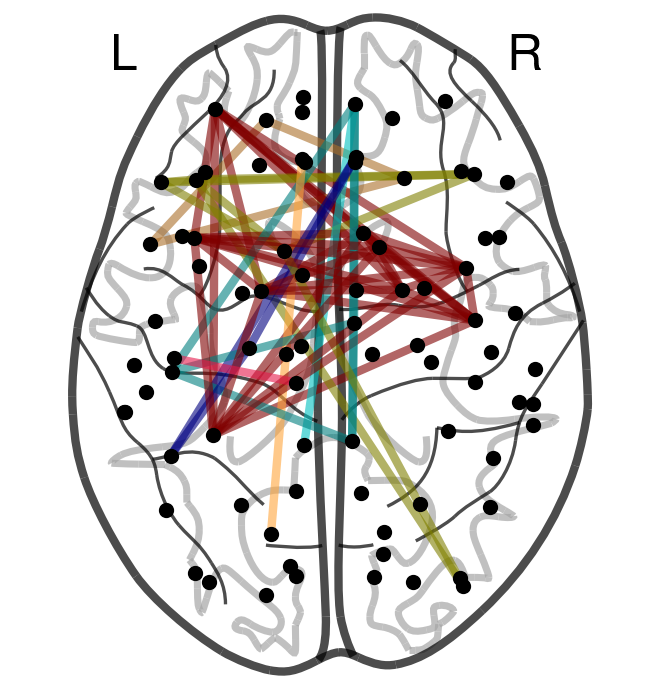}}
\subfloat[Resting State of Subject 3]{\includegraphics[width=0.3\textwidth]{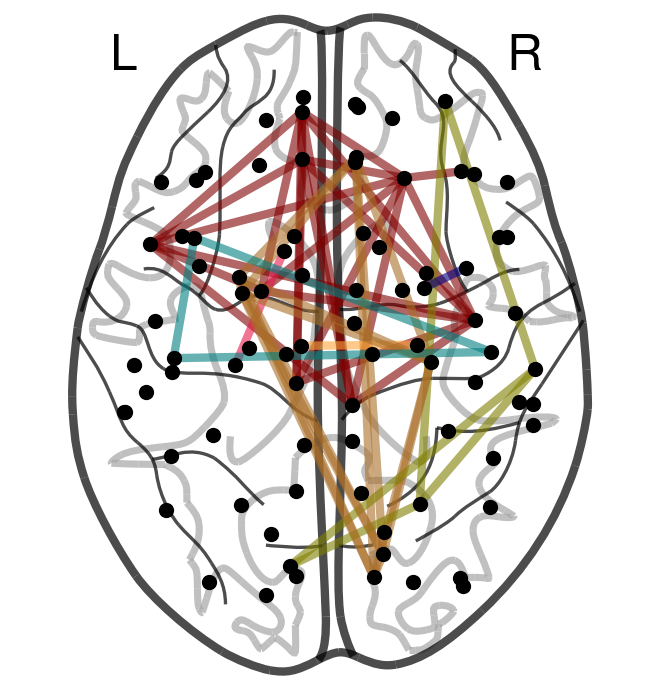}}\\
\subfloat[Language Task of Subject 1]{\includegraphics[width=0.3\textwidth]{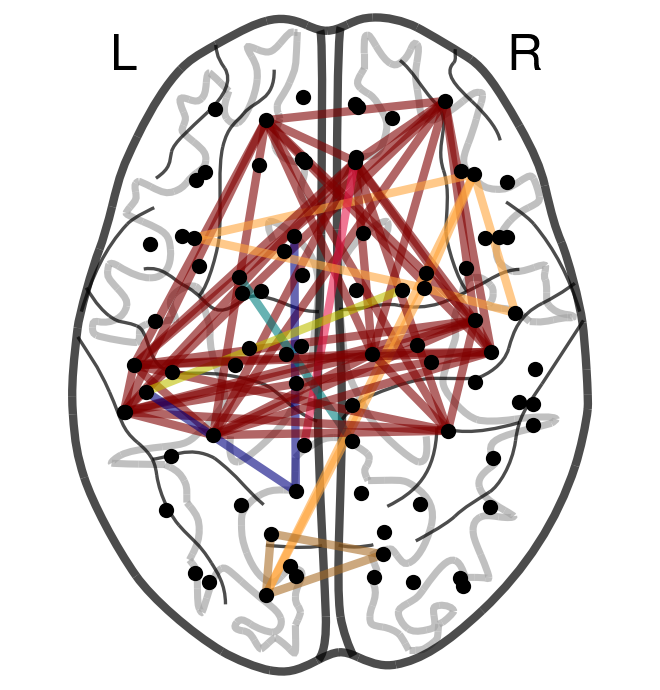}}
\subfloat[Language Task of Subject 2]{\includegraphics[width=0.3\textwidth]{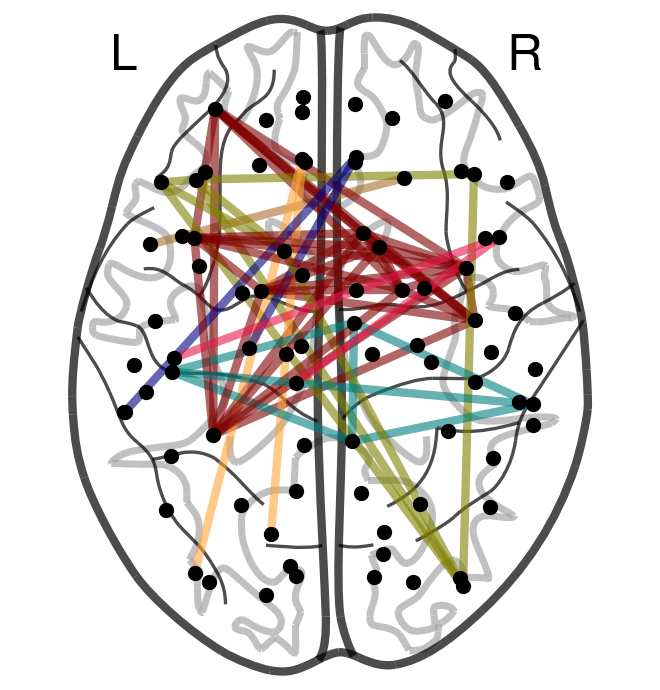}}
\subfloat[Language Task of Subject 3]{\includegraphics[width=0.3\textwidth]{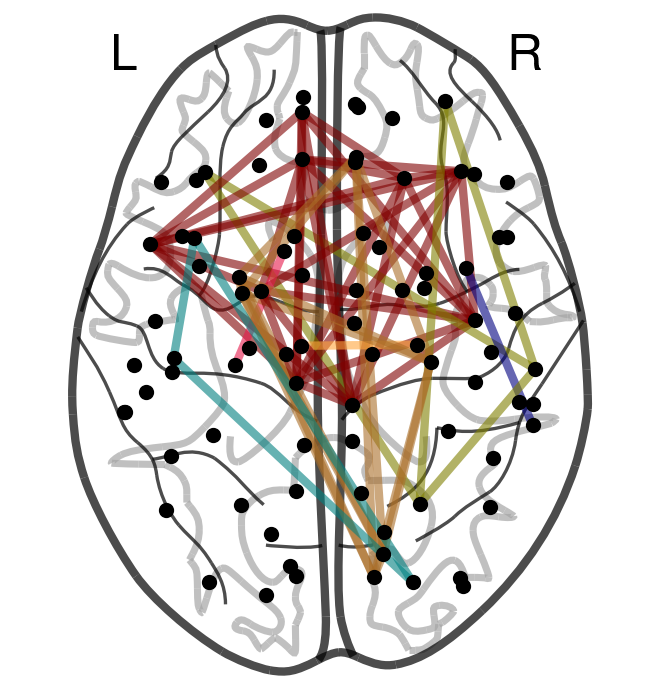}}\\
\subfloat{\includegraphics[width=\textwidth]{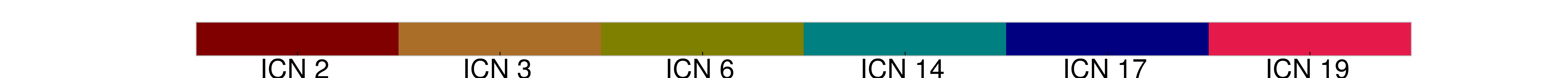}}
\caption{Individual ICNs with expected \gls{icn} amplitude, $\exp\left(\mu_{k,s}^g\right)$, of 3 different subjects under resting state and language task. We only show the edges whose corresponding functional connectivities are not smaller than the 99\% quantile for each \gls{icn}. Functional connectivity is measured by correlation. Edges in different colors belong to different individual ICNs.}
\label{fig:ICNind}
\end{figure}

\Cref{fig:ICNind} illustrates how a set of \glspl{icn} contribute to a dynamic functional brain network. Each row shows the individual difference of the same \gls{icn} under the same experimental condition. Each column shows the changes of individual \glspl{icn} contribute to a similar yet different dynamic functional connectivity. This analysis demonstrates the advantage of \gls{bicnet} over group \gls{ica} methods because it provides a framework upon which we infer the \glspl{icn} with dynamic amplitudes other than just spatial maps. 

\begin{figure}
\centering
\includegraphics[width=.9\textwidth]{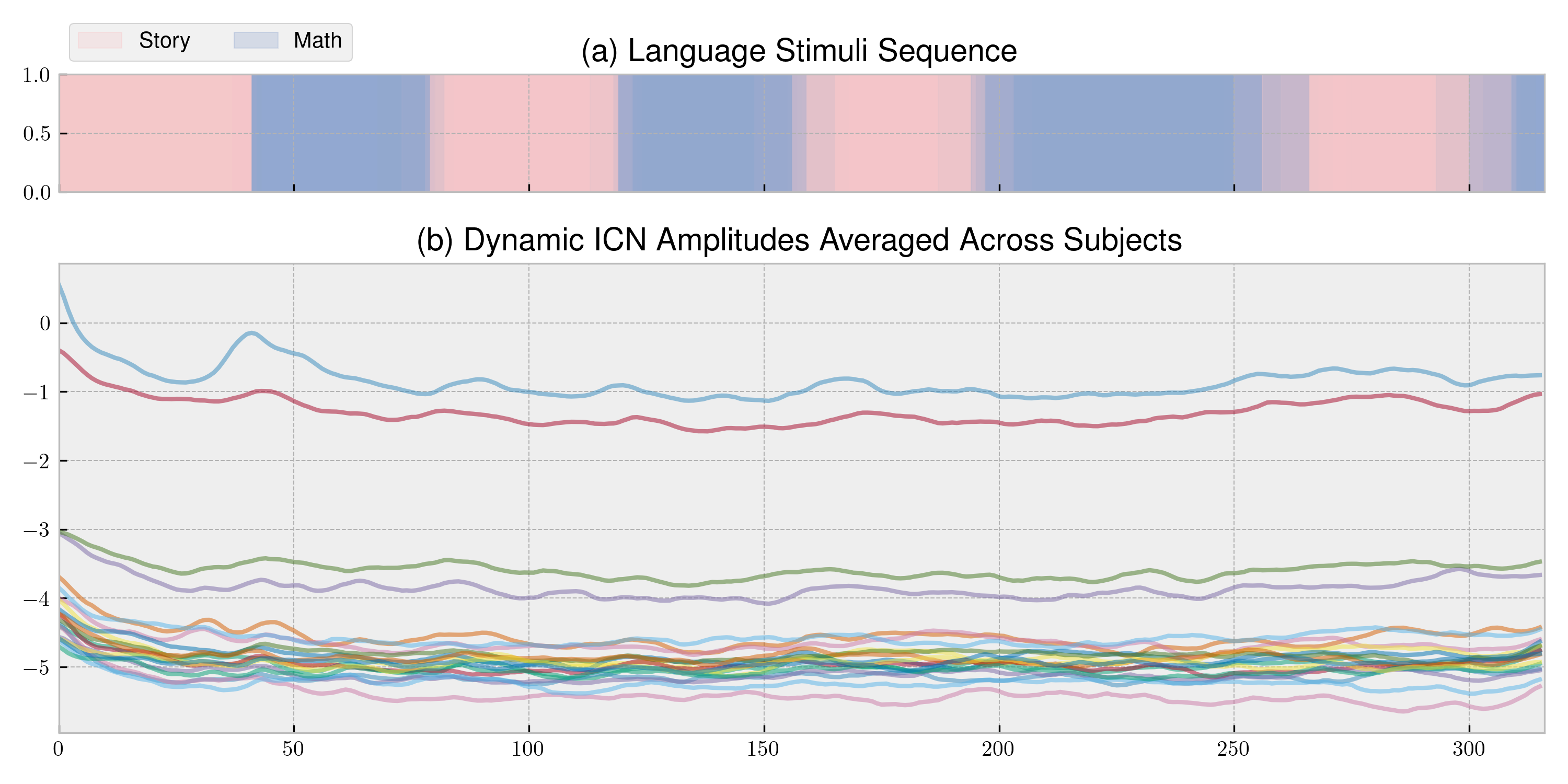}
\caption{(a) Stimuli sequence of the language task. Each subject was required to finish 4 story tasks and 4 math tasks. The duration of each task can be different. (b) The dynamic \gls{icn} amplitudes averaged across subjects.}
\label{fig:dynampaver}
\end{figure}

In \gls{bicnet}, the dynamic amplitude is measured by $\exp\left(h_{k,t,s}^g\right)$. In this data analysis, we label the resting period and language task period a priori by assigning the corresponding task $g$ to each time point $t$. During the language task period, the participants were given either story or math tasks whose sequence is unknown to the \gls{bicnet} a priori, as shown in \cref{fig:dynampaver}(a.

To interpret the dynamic amplitudes, we compared these to the stimulus sequence in \cref{fig:dynampaver}(a) using both time-lagged cross-correlation and change point detection. First, we calculate the time-lagged cross-correlation with lag $\Delta t\in\{1,\dots,50\}$ between stimulus sequence and the dynamic amplitudes for each subject and each \gls{icn}. The time lag of maximum cross-correlation is at $17.28$ seconds with a standard deviation of $12$ seconds. 

\begin{figure}[htbp]
\centering
\subfloat{\includegraphics[width=.9\textwidth]{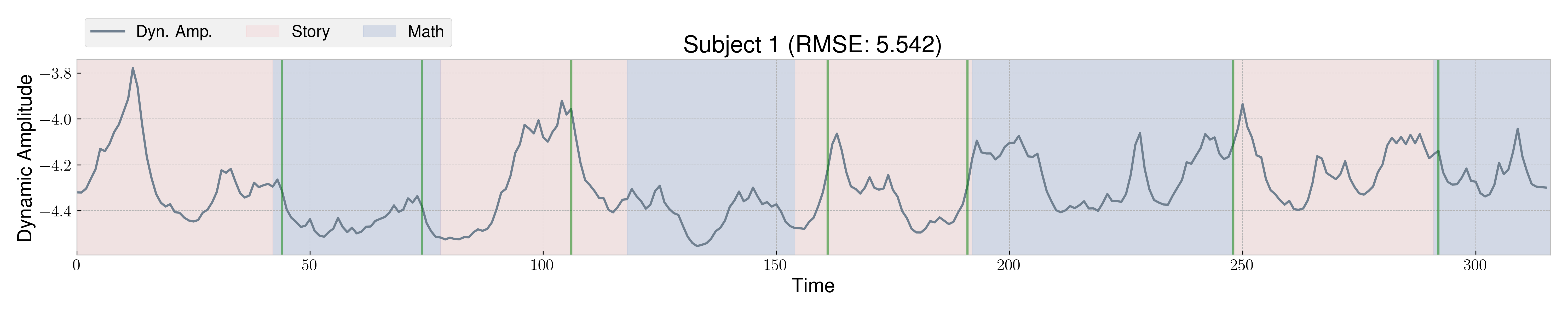}}\\
\subfloat{\includegraphics[width=.9\textwidth]{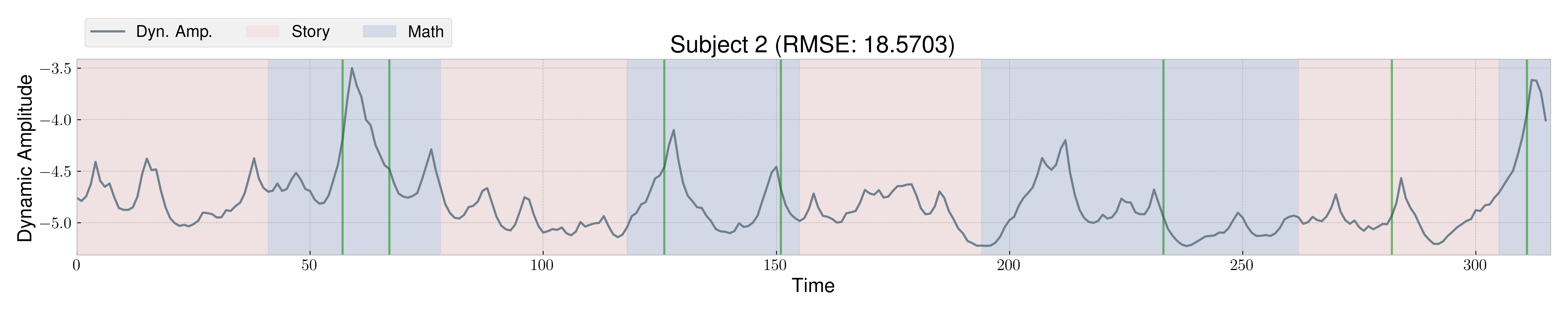}}\\
\subfloat{\includegraphics[width=.9\textwidth]{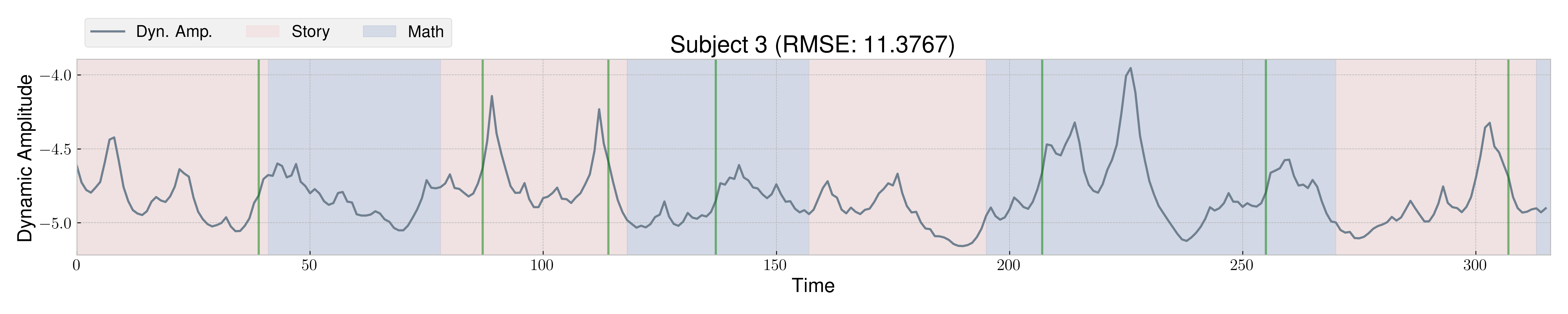}}\\
\subfloat{\includegraphics[width=.9\textwidth]{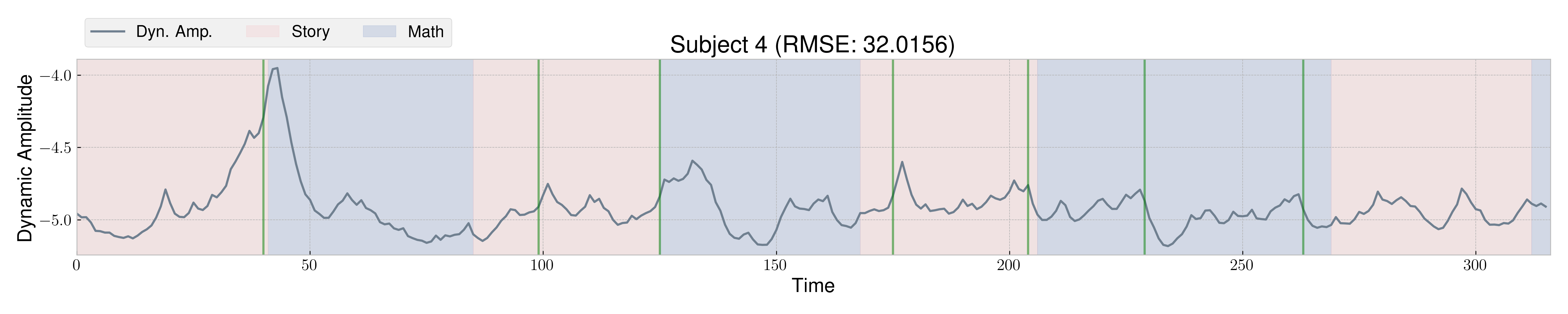}}\\
\subfloat{\includegraphics[width=.9\textwidth]{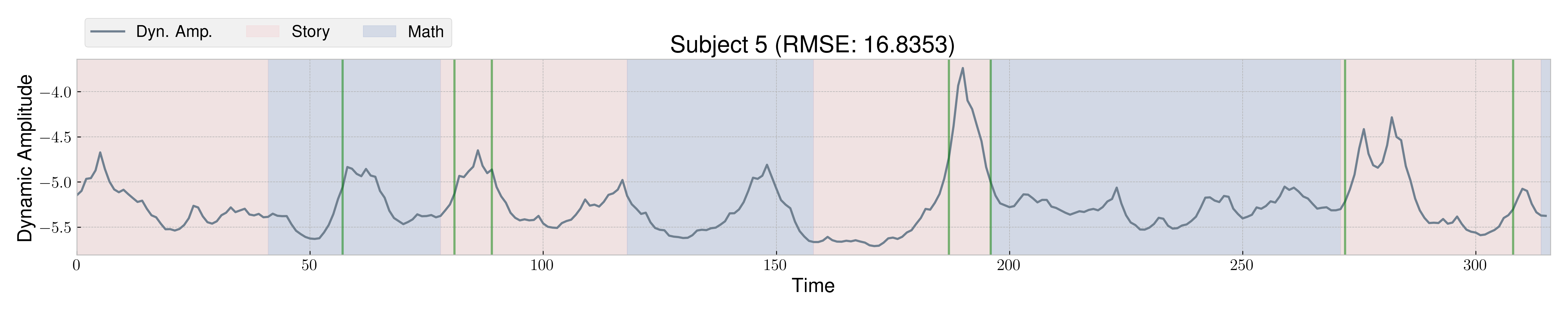}}
\caption{Dark blue curves are dynamic \gls{icn} amplitudes of the first 5 individuals for the same \gls{icn}. Green vertical lines are the detected change points that are closest the detected by Bayesian change point analysis with higher than $70\%$ posterior probability. As shown in the subtitle of each figure, we also calculated the \gls{rmse} between the detected change points in green and the actual stimuli sequence.}
\label{fig:TsCorrCp}
\end{figure}

As shown in \cref{fig:TsCorrCp}, a change-point analysis was conducted using the Bayesian method in \citep{Erdman2007a} where the \gls{rmse} was calculated. The Bayesian change point algorithm tends to detect more change points from the dynamic amplitudes than the real change of stimuli. The 25\%, 50\%, and 75\% quantiles of \glspl{rmse} are 8.87 seconds, 15.58 seconds, and 30 seconds respectively, which is highly similar to the time lag estimated by the time-lagged cross-correlation above. 

The results suggest a time lag for the dynamic amplitudes to adapt to the stimulus changes, and the dynamic amplitudes fluctuate more frequently than the external stimuli. 

\section{Conclusion}
The proposed \gls{bicnet} model was demonstrated to have the ability to identify individual and group-level \gls{icn} structures and quantify the task-related effect through changes in \gls{icn} amplitudes in the single-subject level. The hierarchy \gls{icn} structure is flexible enough to account for individual differences while assuming a group-level structure. The sparsity provides an interpretable \gls{icn} structure. The simulation studies demonstrate that \gls{bicnet} is a potentially more informative substitute to \gls{ica}-based methods on quantifying task-related effects on a set of \glspl{icn}. In the analysis of \gls{hcp} data, \gls{bicnet} identified language networks across subjects and limbic networks related to emotional processing and showed high inter-subject variability. It also suggests that language task execution involves language comprehension, attention, and movement execution.

The stochastic volatility process on the \gls{icn} amplitudes assumes continuous and nonlinear brain states. Similar to \gls{ica}, \gls{bicnet} can handle various types of non-Gaussianity with the stochastic volatility processes, such as sub-Gaussian and super-Gaussian distributions, and can be adapted to skewed distributions. However, \gls{bicnet} does not admit multi-modal distributions because each \gls{icn} has a distinct singular functional role, which should not be mistaken for functional segregation regions of interest.

Determining the number of factors is a tricky question in latent factor modeling. Under the Bayesian framework, there are many ways to tackle the problem. One approach is to estimate the model with different $K$'s and then compare the samples using model selection criteria. However, this method is time- and resource-consuming and not applicable to a large-scale and complicated real-life problem. The second option is by using sparse priors that are robust against over-fitting on $K$. For example, we can specify a large $K$ and let the sparse prior shrink the superfluous factor loadings to zero. However, this method's robustness is yet to confirm and is also not scalable to a high-dimensional problem. The third is to adopt different MCMC sampling algorithms, such as reversible jump MCMC to explore possible $K$. It is also a common practice in Bayesian factor analysis. \cite{Conti2014} applied this algorithm to the Bayesian explanatory factor model and inferred the number of latent factors. More advance, we can use Bayesian nonparametric to model the unknown dimension of latent space. In terms of factor analysis, the Indian buffet process is the most commonly used \cite{Knowles2011}.

The use of spike-and-slab prior gives satisfying results in terms of interpretability. However, the MCMC sampling methods tend to mix slowly due to the posterior distribution's multi-modal nature. Also, the hidden layers in \gls{bicnet} are computationally challenging. This issue can be alleviated by using approximation inference methods, including variants of variational Bayes approximations, variants of expectation propagation methods \cite{Stegle2000}, and integrated nested Laplace approximations \cite{Rue2009}. 

Further, in the analysis, the experimental condition $g\in\{1,\dots,G\}$ is known in advance. In reality, we sometimes do not know the experimental conditions, but we are interested in inferring them. \cite{Ting2019} proposes to use a Markov-switching model to estimate unknown stimuli sequence and apply it to the \gls{hcp} language and motor tasks. However, it requires the number of experimental conditions known or estimated with model selection methods in advance. Besides, \cite{Fox2010} proposes a Bayesian nonparametric approach for Markov-switching processes, which can also infer the number of experimental conditions. However, a Markov-switching model will add one more hidden layer to the \gls{bicnet} model, making the inference more challenging. Also, Bayesian nonparametric is computationally demanding. Therefore, it is still an open question on how to infer the experimental conditions systematically and effectively.

\bibliographystyle{elsarticle-num}
\bibliography{library}

\end{document}